\documentclass[12pt]{article}
\usepackage{amsmath,amssymb}
\usepackage{latexsym,graphicx,color,bbm}

\def\nn{\nonumber\\}

\def\c{{\rm c}}

\def\6#1{{\underline{#1}}}
\def\m6#1{{\underline{#1}\,}}

\newdimen\Tdim
\def\ispan{{\setbox0=\hbox{i}%
\Tdim\ht0\advance\Tdim\dp0\rule[-\dp0]{0pt}{\Tdim}}}
\def\jspan{{\setbox0=\hbox{j}%
\Tdim\ht0\advance\Tdim\dp0\rule[-\dp0]{0pt}{\Tdim}}}
\def\Tspan#1{{\setbox0=\hbox{#1}%
\Tdim\ht0\advance\Tdim\dp0\advance\Tdim.55ex\rule[-\dp0]{0pt}{\Tdim}\box0}}

\def\be{\begin{eqnarray}}
\def\ben{\begin{eqnarray*}}
\def\ee{\end{eqnarray}}
\def\een{\end{eqnarray*}}
\def\Tr{{\rm Tr}}

\def\p{\partial}

\def\=:{=\hspace{-.7em}\raisebox{1.1ex}{.}\hspace{.1em}\raisebox{-0.2ex}{.} }

\newcommand{\NF}{{N_{\rm F}}}
\newcommand{\MF}{{M_{\rm F}}}
\newcommand{\NC}{{N_{\rm C}}}
\newcommand{\MC}{{M_{\rm C}}}
\newcommand{\hs}[1]{\hspace{#1 mm}}

\newcommand {\beq}{\begin{eqnarray}}
\newcommand {\eeq}{\end{eqnarray}}

\makeatletter
\@addtoreset{equation}{section}

\renewcommand{\thefootnote}{\fnsymbol{footnote}}
\newcommand{\thetablename}{Table}
\def\fnum@table{\thetablename\ \thetable}
\makeatother

\begin{document}
\thispagestyle{empty}
\begin{flushright}
DAMTP-2008-82\\
IFUP-TH/2008-22\\
TIT/HEP 589\\
September, 2008 \\
\end{flushright}
\begin{center}
{\large \bf $SO$ and $USp$ K\"ahler and Hyper-K\"ahler Quotients and
  Lumps} \\ 
\vspace{5mm}

Minoru~Eto$^{1,2}$
\footnote{e-mail address: minoru(at)df.unipi.it},
Toshiaki~Fujimori$^3$
\footnote{e-mail address: fujimori(at)th.phys.titech.ac.jp},
Sven~Bjarke~Gudnason$^{1,2}$
\footnote{e-mail address: gudnason(at)df.unipi.it},\\
Muneto~Nitta$^4$
\footnote{e-mail address: nitta(at)phys-h.keio.ac.jp} and
Keisuke~Ohashi$^5$
\footnote{e-mail address: K.Ohashi(at)damtp.cam.ac.uk}

\bigskip
{\it
$^1$  
~Department of Physics, University of Pisa,
Largo Pontecorvo, 3, Ed. C, 56127 Pisa, Italy
\\
$^2$  INFN, Sezione di Pisa,
Largo Pontecorvo, 3, Ed. C, 56127 Pisa, Italy \\
$^3$ Department of Physics, Tokyo Institute of
Technology, Tokyo 152-8551, Japan\\
$^4$ Department of Physics, Keio University, Hiyoshi,
Yokohama, Kanagawa 223-8521, Japan\\
$^5$ Department of Applied Mathematics and Theoretical Physics,
University of Cambridge, CB3 0WA, UK\\
}

\vspace{5mm}

\abstract{We study non-linear $\sigma$ models whose target spaces
  are the Higgs phases of supersymmetric $SO$ and $USp$ gauge theories
  by using the K\"ahler and hyper-K\"ahler quotient constructions. We
  obtain the explicit K\"ahler potentials and develop an expansion
  formula to make use of the obtained potentials from which we also
  calculate the curvatures of the manifolds. 
  The 1/2 BPS lumps in the $U(1)\times SO$ and $U(1)\times USp$
  K\"ahler quotients and their effective descriptions are also
  studied. In this connection, a general relation between moduli
  spaces of vortices and lumps is discussed. We find a new singular
  limit of the lumps with non-vanishing sizes in addition to the
  ordinary small lump singularity. The former is due to the existence of
  singular submanifolds in the target spaces. }

\end{center}

\vfill
\newpage
\setcounter{page}{1}
\setcounter{footnote}{0}
\renewcommand{\thefootnote}{\arabic{footnote}}

\section{Introduction}

The target space of the ${\cal N}=1$ and ${\cal N}=2$ supersymmetric 
non-linear sigma models (NL$\sigma$M), with four and eight supercharges,
must be K\"ahler \cite{Zumino:1979et} 
and hyper-K\"ahler \cite{AlvarezGaume:1981hm}, 
respectively.  
By using this fact, the notion of the hyper-K\"ahler quotient was
first found in physics  
\cite{Curtright:1979yz,Lindstrom:1983rt} 
and was later formulated mathematically
\cite{Hitchin:1986ea}. 
(We recommend Ref.~\cite{Antoniadis:1996ra} as a review for physicists).
A $U(1)$ hyper-K\"ahler quotient \cite{Curtright:1979yz} 
recovers the Calabi metric \cite{Ca} on the cotangent bundle 
over the projective space, $T^* {\mathbb C}P^{N-1}$, 
while its $U(\NC)$ generalization 
leads to the cotangent bundle 
over the complex Grassmann manifold, $T^* G_{N,\NC}$ 
\cite{Lindstrom:1983rt}. 
The hyper-K\"ahler manifolds also appear in the moduli spaces 
of Bogomol'nyi-Prasad-Sommerfield (BPS) solitons such as 
Yang-Mills instantons \cite{Atiyah:1978ri,Christ:1978jy,Dorey:1999pd} and 
BPS monopoles \cite{Atiyah:1985dv}.
The hyper-K\"ahler quotient offers a powerful tool to 
construct these hyper-K\"ahler manifolds:
instanton moduli spaces \cite{Atiyah:1978ri}  
and monopole moduli spaces \cite{Gibbons:1996nt}. 
Gravitational instantons 
\cite{Kronheimer:1989zs,Lindstrom:1999pz}, 
Yang-Mills instantons on gravitational instantons \cite{KN} 
and toric hyper-K\"ahler manifolds \cite{Bielawski:1998bb}
are all constructed using the hyper-K\"ahler quotient.

The Higgs branch of  
${\cal N}=2$ supersymmetric QCD is hyper-K\"ahler.
The low energy effective theory on the Higgs branch  
is described by an ${\cal N}=2$ NL$\sigma$M 
on the hyper-K\"ahler manifold 
\cite{Seiberg:1994aj,Argyres:1996eh,Argyres:1996hc}.
In the cases of an $SU(\NC)$ or a $U(\NC)$ gauge theory with
hypermultiplets charged commonly under $U(1)$, 
the explicit metrics on the Higgs branch 
and their K\"ahler potentials are known explicitly. 
The latter is nothing but the Lindstr\"om-Ro\v{c}ek metric 
\cite{Lindstrom:1983rt}. 
A $U(1) \times U(1)$ gauge theory with three hypermultiplets 
of certain charges for instance
gives the space: $T^* F_n$ with 
$F_n$ being the Hirzebruch surface \cite{Eto:2005wf}.
The Higgs branches of quiver gauge theories are 
gravitational instantons and Yang-Mills instantons on 
gravitational instantons \cite{Kronheimer:1989zs,KN}.
However, to our knowledge, the ones of an $SO$ or a $USp$ gauge theory  
has not been explicitly derived yet 
(except for $SO(2)\simeq U(1)$ and $USp(2) \simeq SU(2)$),
which was an open question in \cite{Antoniadis:1996ra}.

The first purpose of this paper is to explicitly construct the metric
and its K\"ahler potential on the Higgs branch of  
${\cal N}=1$ and ${\cal N}=2$ supersymmetric gauge theories 
with gauge groups $SO(\NC)$ and $USp(2\MC)$ 
or $U(1)\times SO(\NC)$ and $U(1)\times USp(2\MC)$. 
The vacua of ${\cal N}=1$ supersymmetric gauge theories are determined by 
the $D$-term condition, $D=0$, while those of ${\cal N}=2$ theories
are determined by both
the $D$-term and the $F$-term conditions,
$D=F=0$. 
The moduli space of vacua is obtained by the space of 
solutions to these constraints modulo gauge groups, 
$\left\{ D=0\right\} /G$ 
and $\left\{ D=F=0 \right\} /G$ 
for ${\cal N}=1$ and ${\cal N}=2$ models, respectively.  
In the superfield formalism, solving the $D$-term condition and
modding out the gauge group $G$,
can be done simultaneously because the gauge
symmetry is in fact complexified to $G^{\mathbb C}$. As a bonus the
K\"ahler potentials are directly obtained in the superfield
formalism. 
Although the $D$-term conditions of $SU(\NC)$ and $U(\NC)$ gauge groups
can be solved in components easily, those of $SO(\NC)$ and $USp(2\MC)$ are
difficult to solve.  
To our knowledge this has not been done yet. 
We use the superfield formalism to solve the $D$-term conditions 
for $SO(\NC)$ and $USp(2\MC)$ gauge groups by introducing a trick. 
Namely, we relax the algebra of the vector superfields $V$ from
$\mathfrak{so}(\NC)$ and $\mathfrak{usp}(\NC=2\MC)$ to
$\mathfrak{u}(\NC)$ and then introduce 
a Lagrange multiplier to restrict the algebra of $V$ to
$\mathfrak{so}(\NC)$ and $\mathfrak{usp}(2\MC)$. 
We then successfully solve the superfield equations to obtain the resultant 
K\"ahler potentials.

There exists another method 
to obtain the moduli space of vacua, which is
more familiar in the literature; 
It is an algebro-geometrical method in
the geometric invariant theory \cite{Luty:1995sd},  
in which one prepares holomorphic gauge invariants made of the original 
chiral superfields and looks for algebraic constraints among them. 
This method has been widely used in the studies of ${\cal N}=1$ 
supersymmetric gauge theories
\cite{Intriligator:1995au,Intriligator:1995id,Intriligator:1995ne}. 
See \cite{Gray:2008yu,Hanany:2008kn} for recent developments.
In particular in Ref.~\cite{Hanany:2008kn}, the moduli spaces of vacua of 
${\cal N}=1$ supersymmetric $SO(\NC)$ and $USp(2\MC)$ gauge theories 
are found to be Calabi-Yau cones over certain weighted projective spaces.
According to us, a weak point of the geometric invariant theory is that 
one has to solve algebraic constraints among invariants 
in order to calculate geometric quantities such as the metric and the
curvature etc.

Compared with this situation our method provides the K\"ahler potentials 
directly. We rewrite them in terms of holomorphic gauge invariants. 
Furthermore, we calculate the metrics and the curvatures 
by expanding the K\"ahler potentials.
We confirm that a singularity appears in the moduli space of vacua
when the gauge symmetry is partly recovered, as expected. 
We then study the case of $U(1)\times SO(\NC)$ and $U(1)\times USp(2\MC)$
gauge theories.  
Finally, we calculate $SO(\NC)$ and $USp(2\MC)$ hyper-K\"ahler quotients 
and obtain their K\"ahler potentials explicitly. 
Although only the lowest dimensional case $USp(2) \simeq SU(2)$ has
been known so far \cite{Antoniadis:1996ra}, the higher dimensional cases
are new. 

We find explicitly the K\"ahler quotients for both the ${\cal N}=1$
and some of the ${\cal N}=2$ theories with $SO$ and $USp$ gauge
groups, however, at the classical level. For the ${\cal N}=2$ case we
are in good shape due to the well-known non-renormalization theorem on
the Higgs branch by Argyres-Plesser-Seiberg \cite{Argyres:1996eh},
which leaves the 
results of the metric and K\"ahler potential quantum mechanically
exact. The situation is not quite so good in the ${\cal N}=1$
case. Quantum corrections are to be considered, except in the compact
directions of the Nambu-Goldstone modes (up to overall constants: pion
decay constants) which is indeed consistent with the low-energy
theorem of Nambu-Goldstone modes. Along the non-compact directions
parametrized by quasi-Nambu-Goldstone modes the corrections are out of
control and can render rather large. All in all, the total K\"ahler
potential is correct only (semi-)classically for the ${\cal N}=1$ case
and it will take the form 
\beq K = f(I_1,I_2,\ldots) \ , \eeq
with $I_a$ being $G^{\mathbb{C}}$ invariants and $f$ some function. In
the case of ${\cal N}=1, U(\NC)$ theories, some quantum corrections has
been considered in the literature \cite{Grisaru:1982df}.
To this end, we emphasize that the metric and K\"ahler potential was
until now unknown, even classically and the first step has been taken,
which of course leaves the quantum corrections as an important and
interesting future calculation to grasp.

The second purpose of this paper is concerned with sigma model lumps, 
or sigma model instantons.
A lump solution was first found in the $O(3)$ sigma model, 
or the ${\mathbb C}P^1$ model \cite{Polyakov:1975yp}.
It was then generalized to the ${\mathbb C}P^n$ model \cite{Din:1980jg},
the Grassmann model \cite{Din:1981bx}, 
and other K\"ahler coset spaces \cite{Perelomov:1987va}.  
Lumps are topological solitons associated with 
$\pi_2(M)$ with $M$ being the target K\"ahler manifold.
Their energy saturates the BPS bound of 
the topological charge written 
as the K\"ahler form of $M$ pulled-back to 
the two-dimensional space.\footnote{
In the case of hyper-K\"ahler manifolds 
there exist triplets of complex structures and K\"ahler forms.
Accordingly it has recently been found that 
there exists a BPS bound written by the sum of 
three different K\"ahler forms to three different planes 
in the three dimensional space 
\cite{Naganuma:2001pu}.}
The lump solutions preserve half of supersymmetry, when embedded into
supersymmetric theories. 
The dynamics of lumps was studied \cite{Ward:1985ij}
by the moduli space (geodesic) approximation. 
Lumps are related to vortices in gauge theories as follows. 
$U(1)$ gauge theories coupled to several Higgs fields 
often admit semi-local vortex-strings \cite{Vachaspati:1991dz}.  
In the strong gauge coupling limit, 
gauge theories reduce to NL$\sigma$Ms 
whose target space is the moduli space of vacua in 
the gauge theories, 
and in this limit, semi-local
strings reduce to lump-strings. 
For instance, a $U(1)$ gauge theory coupled to two 
charged Higgs fields reduces to the ${\mathbb C}P^1$ model, 
while the semi-local vortex-strings in Ref.~\cite{Vachaspati:1991dz}
reduce to the ${\mathbb C}P^1$ lumps \cite{Hindmarsh:1991jq}.
In the gauge theories at finite coupling, 
the large distance behavior of semi-local strings 
is well approximated by lump solutions. 
The sizes or widths of semi-local strings 
are moduli of the solution in the BPS limit, and 
accordingly, the lumps also possess size moduli.
When the size modulus of a semi-local string vanishes,
the solution reduces to the Abrikosov-Nielsen-Olesen (ANO) vortex
\cite{Abrikosov:1956sx} which is called a local vortex. This limit
corresponds to a singular configuration in the NL$\sigma$M, which is
called the small lump singularity.
Lumps and semi-local strings are also candidates of cosmic strings,
see e.g.~Ref.~\cite{Benson:1993at}, and appear also in recent studies of
D-brane inflation etc.~\cite{Dasgupta:2007ds}.

Recently, there has been much progress on non-Abelian vortices 
in $U(\NC)$ gauge theories \cite{Hanany:2003hp,Auzzi:2003fs}. These
vortices are naturally 1/2-BPS in ${\cal N}=2$ supersymmetric
theories.  
When the number of flavors $\NF$ is equal to 
the number of colors $\NC$, 
the theory admits local non-Abelian vortices. 
Each of them carries orientational moduli ${\mathbb C}P^{\NC-1}$ 
in the internal space. 
The determination of the full moduli space of 
multiple local vortices with arbitrary positions 
and arbitrary orientations were achieved in field theory 
\cite{Eto:2005yh} by introducing the method of 
the ``moduli matrix'' \cite{Isozumi:2004vg,Eto:2006pg}. 
All the moduli parameters are contained 
in the moduli matrix, which is a holomorphic matrix of 
the same size as the Higgs fields, 
and the moduli space has been shown to 
coincide with the one \cite{Hanany:2003hp,Hashimoto:2005hi}
conjectured in string theory. 
The dynamics of two non-Abelian vortices has been 
studied in the moduli space approximation \cite{Eto:2006db} by using
the general formula for the effective action of BPS solitons
\cite{Eto:2006uw}. 
Many interesting aspects of non-Abelian vortices 
are reviewed in
Refs.~\cite{Eto:2006pg,Tong:2005un,Konishi:2007dn,Shifman:2007ce}.
For instance, monopoles (Yang-Mills instantons) become kinks
\cite{Tong:2003pz,Hanany:2004ea} 
($\mathbb{C}P^{\NC -1}$-lumps \cite{Hanany:2004ea,Eto:2004rz})
in the effective field theory of a vortex-string. 
Intriguing is also the flux matching between non-Abelian vortices and
non-Abelian monopoles and the applications are very interesting in the
connection with non-Abelian duality
etc.~\cite{Auzzi:2003em,Konishi:2007dn}. 
Furthermore, non-Abelian vortices in ${\cal N}=1$ supersymmetric
theories have been studied in Refs.~\cite{Shifman:2005st}. 
A dyonic extension of non-Abelian vortices has been studied recently
in Ref.~\cite{Collie:2008za}. 

In the case of a $U(\NC)$ gauge theory, 
semi-local vortices exist when the number  
of flavors $\NF$ is larger than the number of colors $\NC$
\cite{Shifman:2006kd}. 
At strong gauge coupling, 
the $U(\NC)$ gauge theory reduces to 
the Grassmann sigma model 
on $Gr_{\NF,\NC} = SU(\NF)/[SU(\NF-\NC)\times 
SU(\NC)\times U(1)]$. 
It has been demonstrated in Ref.~\cite{Eto:2007yv} that 
non-Abelian semi-local strings in a $U(\NC)$ gauge theory 
reduce to the Grassmann lumps at large distance. 
One interesting aspect of these lumps (semi-local vortices) is the
(non$\textrm{-)}$normalizability of zero modes. 
It has been shown in Ref.~\cite{Shifman:2006kd} that 
all moduli parameters of a single lump are non-normalizable 
except for its position moduli. 
Orientational moduli in the internal space for local vortices 
are in fact non-normalizable in this case. 
However, in the limit of vanishing size modulus, 
normalizable orientational zero modes appear \cite{Eto:2007yv}. 
More interestingly, for $k=2$ lumps (semi-local vortices), 
their ``relative" orientational moduli are normalizable 
although their ``overall" orientational moduli are non-normalizable 
\cite{Eto:2006db,Eto:2007yv}.

After the discovery of the $U(\NC)$ non-Abelian vortices 
\cite{Hanany:2003hp,Auzzi:2003fs}, 
one remarkable new development is an extension 
to vortices in $U(1)\times SO(\NC)$ gauge theories \cite{Ferretti:2007rp} 
and $U(1)\times G'$ gauge theories with an arbitrary 
simple group $G'$ \cite{Eto:2008yi}. 
This was done by imposing $G'$ invariant constraints 
on the moduli matrix, 
and the conditions for the local vortices in these theories 
have been found. 
In this paper we focus on BPS lumps related to 
semi-local vortices in the $U(1)\times SO(\NC)$ and 
$U(1)\times USp(2\MC)$ gauge theories, 
which is the second purpose of this paper. 
We make a connection between the lump moduli spaces and the vortex
moduli spaces and on this course, introduce the moduli matrix, in which
we have the formalism to explicitly construct $1/2$ BPS lumps in the
class of $U(1)\times G'$ gauge theories. The explicit examples we make
are with $U(1)\times SO(2\MC)$ and $U(1)\times USp(4)$.  
Interestingly, there is a crucial difference between the $U(\NC)$
and $U(1)\times SO(\NC)$ or $U(1)\times USp(2\MC)$ theories, which is
that in the latter two, even for $\NF = \NC$, semi-local vortex strings
appear which is not the case for $U(\NC)$. 

We examine the (non-)normalizability of the moduli parameters of lumps
in the $U(1)\times SO(\NC)$ and $U(1)\times USp(2\MC)$ K\"ahler
quotients.  
In the case of a single lump solution, all moduli parameters in both
the models are non-normalizable except for the center of mass.  
This is parallel to the case of the $U(\NC)$ K\"ahler quotient
\cite{Shifman:2006kd,Eto:2007yv}.

This paper is organized as follows. In Sec.~\ref{sec:review} we will
make a short review on the $SU(\NC)$ and $U(\NC)$ K\"ahler quotients and
also the $U(\NC)$ hyper-K\"ahler quotient while we will turn our attention
to the $SO(\NC)$ and $USp(2\MC)$ and also $U(1)\times SO(\NC)$ and
$U(1)\times USp(2\MC)$ K\"ahler quotients in
Sec.~\ref{sec:kahlerquotient}, 
furthermore construct the metrics, an expansion of the metric around
their vacuum expectation values and compute the corresponding
curvatures. Then we 
make use of the technology with some explicit examples. Finally, we
lift the construction to the hyper-K\"ahler quotient case of $SO(\NC)$
and $USp(2\MC)$ gauge theories. In Sec.~\ref{sec:nlsmlumps} will
consider the NL$\sigma$M lumps, first by general considerations of
gauge theories with $U(1)\times G'$ gauge groups with $G'$ being an
arbitrary simple group. Then we make a connection between the moduli
spaces of the lumps in these theories with the moduli spaces of the
vortices. Finally, we construct the lumps with the target spaces which
we constructed in Sec.~\ref{sec:kahlerquotient}, make effective
descriptions of those, and identify the non-normalizable modes. In
Sec.~\ref{sec:concdisc} we conclude and discuss further developments. 
Moreover, we have left various theorems and proofs used in the text
for Appendix \ref{appvartheorem}, a uniqueness proof in Appendix
\ref{appuniqueness} and a deformed K\"ahler potential for $USp(2\MC)$
in Appendix \ref{appuspdeform}. 

\section{The $SU(\NC)$ and $U(\NC)$ (Hyper-)K\"ahler Quotients: \\
A Review \label{sec:review}}
\subsection{The $SU(\NC)$ and $U(\NC)$ K\"ahler Quotients}

Let us first give a brief review on the $SU(\NC)$ K\"ahler quotient.
We start with the ${\cal N}=1$ $SU(\NC)$ supersymmetric Yang-Mills theory with 
$\NF$ chiral superfields $Q$ (i.e.~an $\NC$-by-$\NF$ matrix)
in the fundamental representation of $SU(\NC)$. Denote the $SU(\NC)$ 
vector multiplet by a superfield $V'$, then a K\"ahler potential for
the system is 
\beq
K_{SU(\NC)} = \Tr\left[QQ^\dagger e^{-V'}\right] \ .
\label{eq:Kahler_SU}
\eeq
We have used a matrix notation and the trace is taken over the color indices.
The Lagrangian is invariant under 
the complexification of the gauge group, 
$SU(\NC)^{\mathbb C} = SL(\NC,{\mathbb C})$, given by 
\beq
Q \to e^{i\Lambda'} Q\ ,\quad
e^{V'} \to e^{i\Lambda'} e^{V'} e^{- i\Lambda'{}^\dagger}\ ,\qquad
e^{i\Lambda'} \in SU(\NC)^{\mathbb C} \ .
\eeq
We do not consider any superpotentials here.

As is well-known, the kinetic term of the vector supermultiplet 
$\int d^2\theta\ W^\alpha W_\alpha /4g^2 + {\rm c.c.}$
includes a so-called $D$-term potential in the Wess-Zumino gauge, 
in which $SU(\NC)^{\mathbb C}$ is fixed to $SU(\NC)$
\beq
 V_D = \frac{g^2}{2}\left(D^{A}\right)^2\ ,\qquad
 D^A = \Tr_{\rm F} \left(Q_{\rm wz}^\dagger T^A Q_{\rm wz}\right)\ ,
\eeq
where $T^A$ are $SU(\NC)$ generators and $Q_{\rm wz}$ is $Q$ in 
the Wess-Zumino gauge. 
The vacuum condition $D^A=0$ ($D$-flatness) allows both for an
unbroken phase and the Higgs phase. 
It implies that 
$Q_{\rm wz}Q^\dagger_{\rm wz} \propto {\bf 1}_\NC$  
holds in the vacuum.
On the Higgs branch (${\rm rank}\,Q_{\rm wz}=\NC$), the gauge fields
acquire masses of the order $g\left<Q\right>$ by the Higgs mechanism. 
If we restrict ourselves to energies much below the mass
scale, we can omit the massive gauge fields. In order to get a low
energy effective theory, it will prove useful to consider a limit
where the gauge coupling is taken to infinity: $g\to\infty$. In this
limit, the vector multiplet becomes infinitely massive and looses the
kinetic term. Thus, it reduces to merely an auxiliary field. At the
same time the $D$-term potential forces $Q_{\rm wz}$ to take a value
in the vacuum $D^A=0$. 
Thus, the low energy effective theory is a non-linear sigma model
(NL$\sigma$M), whose target space is the vacuum of 
the gauge theory 
\beq
{\cal M}_{SU(\NC)} = \left\{Q_{\rm wz}\ |\ 
Q_{\rm wz}Q^\dagger_{\rm wz} \propto {\bf 1}_\NC\ ,\, 
{\rm rank}\,Q_{\rm wz}=\NC\right\}/SU(\NC) \ .
\label{eq:target_D}
\eeq
The real dimension of the manifold is $2 \NC\NF - (\NC^2-1) - (\NC^2-1) =
2\NC(\NF-\NC)+2$. 

Before fixing the complexified gauge symmetry $SU(\NC)^{\mathbb C}$, for
example by the Wess-Zumino gauge as above, we can take the strong
coupling limit. This gives another description of the non-linear sigma
model. The Lagrangian consists of only one term
i.e.~Eq.~(\ref{eq:Kahler_SU}).
We do not have the $D$-term conditions anymore, however, instead we
have the complex fields $Q$ and the complexified gauge group
$SU(\NC)^{\mathbb C}$. 
The target space is expressed by
\beq
{\cal M}_{SU(\NC)} = \{Q \, | \, {\rm rank} \, Q = \NC \}/\!\!/SU(\NC)^{\mathbb C}\ .
\label{eq:target_C}
\eeq
In order for this quotient to be well-defined, the action of
$SU(\NC)^{\mathbb C}$ must be free on $Q$. 
Namely, the gauge symmetry should be completely broken, thus we are
going to study the {\it full} Higgs phase.  
The complex dimension of the manifold is $\NC\NF - (\NC^2-1) =
\NC(\NF-\NC)+1$, which coincides with the dimension of (\ref{eq:target_D}).
The two expressions (\ref{eq:target_D}) and (\ref{eq:target_C}) of the
target space are identical. 
One can find a relation between them by solving
the equations of motion for $V'$. It determines the traceless part as
$QQ^\dagger e^{-V'} \propto {\bf 1}_\NC$.
Taking ${\rm Tr} \, V'=0$ into account, $V'$ is uniquely determined as
\beq
V' =\log QQ^\dagger-\frac1\NC{\bf 1}_\NC \log \det (QQ^\dagger)\ ,
\label{eq:sol_V'_SU}
\eeq
if and only if ${\rm rank}\,Q$ is the maximum, which means the full
Higgs phase.
Then we find an explicit map from the quotient (\ref{eq:target_C}) to
the vacuum configuration (\ref{eq:target_D}):
\beq
Q_{\rm wz} = e^{-V'/2} Q =\left[ \det (QQ^\dagger)
\right]^{\frac{1}{2\NC}} \frac{1}{\sqrt{QQ^\dagger}} Q\ .
\eeq

There exists still another way to express the same NL$\sigma$M. 
As explained above, the target space is nothing but the
classical moduli space of vacua of the original supersymmetric gauge
theory. 
As discussed in Ref.~\cite{Luty:1995sd} 
it can be described by holomorphic invariants of the
complexified gauge group. Hence, the K\"ahler potential on the
NL$\sigma$M should be expressed in terms of such
holomorphic invariants.
The holomorphic invariants of $SU(\NC)^{\mathbb C}$ are the baryon
operators 
\beq
B^{\left<A_1\cdots A_\NC\right>} \equiv \det Q^{\left<A_1\cdots
  A_\NC\right>} = \epsilon^{i_1 \cdots i_\NC} Q_{i_1}{}^{A_1} \cdots Q_{i_\NC}{}^{A_\NC},
\eeq
where $Q^{\left<A_1\cdots A_\NC\right>}$ denotes an $\NC$-by-$\NC$ minor matrix of $\NC$-by-$\NF$ matrix $Q$ as $(Q^{\left<A\right>})_i^j=Q_i{}^{A_j}$.
We often abbreviate the label $\left<A_1\cdots A_\NC \right>$ as
$\left<A\right>$. 
The important point is that all the $B^{\left<A\right>}$'s 
are not independent and they satisfy the so-called
Pl\"ucker relations
\begin{eqnarray}
 B^{\langle A_1\cdots A_{\NC-1}[B_1\rangle }B^{\langle B_2\cdots B_{\NC+1}]
\rangle}=0 \ . \label{eq:Plucker}
\end{eqnarray}
Furthermore, the condition for having the full Higgs phase requires 
that at least one of the $B^{\langle A \rangle}$'s must take a
non-zero value. Actually, we can reconstruct $Q$ modulo $SU(\NC)$ gauge
symmetry by solving the Pl\"ucker relations with one non-zero
$B^{\langle A \rangle}$ as the starting point. 
That is, the holomorphic invariants with the Pl\"ucker relations give us
the same information as the two descriptions above. 
Hence, the target space is also expressed as
\beq
{\cal M}_{SU(\NC)} = \left\{B^{\left<A\right>}\ |\ \text{Pl\"ucker\
  relations\, (\ref{eq:Plucker})}\right\}
-\left\{B^{\left<A\right>}=0 \ ,  \forall \langle A \rangle \right\}.
\label{eq:target_I}
\eeq

Let us show the metric on the target space. It can be derived from the
K\"ahler potential (\ref{eq:Kahler_SU}) and is represented by 
\beq
K_{SU(\NC)} 
= \NC \left[ \det (QQ^\dagger) \right]^{\frac{1}{\NC}} 
= \NC \left( \sum_{\left<A\right>} \left| B^{\left<A\right>}\right|^2\right)^{\frac{1}{\NC}}.
\eeq
The appearance of the $\NC$th root reflects the fact that the $U(1)$ charge
of the invariants is $\NC$, as we will see soon.
Notice that the (partial) Coulomb phase ($\det (QQ^\dagger)=0$)
shrinks to a point of the target manifold from the point of view of the
NL$\sigma$M and a trace of this fact is seen as the $\mathbb{Z}_\NC$
conifold singularity at that point. 
In a simple example with $\NF = \NC$, one can find 
the NL$\sigma$M on an orbifold ${\mathbb C}/{\mathbb Z}_\NC$. 
At the singularity, the vector multiplet becomes massless and the
gauge symmetry is restored. We have to take all the massless fields
into account there, namely we cannot restrict ourselves to the
NL$\sigma$M, but we have to return to the original gauge theory.

\medskip
This singularity (that is, the Coulomb phase) 
is removed once the
overall $U(1)$ phase is gauged and the
so-called Fayet-Iliopoulos (FI) parameter $\xi\ (>0)$
\cite{Fayet:1974jb} is introduced for that $U(1)$.
Let us consider a $U(1) \times SU(\NC)$ gauge theory.
Still we neglect the kinetic terms associated with the vector
multiplet, such that the vector multiplet is an auxiliary superfield. 
The K\"ahler potential is given by
\beq
K_{U(1) \times SU(\NC)} = \Tr\left[QQ^\dagger e^{-V_e}e^{-V'}\right] + \xi V_e
= e^{-V_e} K_{SU(\NC)} + \xi V_e\ ,
\label{eq:Kahler_U}
\eeq
where $V_e$ is a $U(1)$ vector supermultiplet and the chiral fields $Q$
have $U(1)$ charge $+1$. 
The $D$-flatness condition for the overall $U(1)$ implies that 
$Q_{\rm wz}Q^\dagger_{\rm wz} = \frac{\xi}{\NC} \mathbf 1_\NC$.
The target space of the NL$\sigma$M becomes a compact space; 
the complex Grassmannian manifold ${\cal M}_{U(1) \times SU(\NC)} =
Gr_{\NF,\NC} \simeq SU(\NF)/[SU(\NF-\NC) \times SU(\NC) \times U(1)]$.\footnote{
The $U(\NC)$ K\"ahler quotient construction of the Grassmann 
manifold was first found in Ref.~\cite{Aoyama:1979zj}
in the superfield formalism.
}
As in the case above, we have three different representations
\begin{align}
{\cal M}_{U(1) \times SU(\NC)} &= \left\{Q_{\rm wz}~\Big|~
Q_{\rm wz} Q_{\rm wz}^\dagger = \frac{\xi}{\NC} {\bf 1}_\NC
\right\}\bigg/\left(U(1) \times SU(\NC)\right) \nonumber\\
&= \left\{ Q \, | \, {\rm rank} \, Q = \NC \right\}/\!\!/(U(1) \times
SU(\NC))^{\mathbb C} \nonumber\\ 
&= \left( \left\{B^{\left<A\right>}\ |\ \text{Pl\"ucker
  relations \, (\ref{eq:Plucker})}\right\} - \left\{B^{\left<A\right>}=0 \ ,  \forall \langle A \rangle \right\} \right)/\!\!/U(1)^{\mathbb C}\ .
\label{eq:grassmaniann}
\end{align}
A relation between $Q_{\rm wz}$ and $Q$ is also found here by solving
the equations of motion with respect to $V'$ and $V_e$. 
The solution for $V'$ is the same as Eq.~(\ref{eq:sol_V'_SU}) 
and the $U(1)$ part is then written as
\begin{eqnarray}
 V_e=\log\left(\xi^{-1} K_{SU(\NC)}\right) \ .
\end{eqnarray}
Then the map from the quotient space to the vacuum configuration is
given by
\begin{eqnarray}
 Q_{\rm wz} = e^{-V'/2-V_e/2} Q =
 \sqrt{\frac{\xi}{\NC}}  \frac{1}{\sqrt{QQ^\dagger}} Q \ . 
\end{eqnarray}
The third expression in Eq.~(\ref{eq:grassmaniann}) 
shows the Pl\"ucker embedding of the Grassmannian space into a bigger
space, the complex projective space ${\mathbb C}P^{n}$ with $n = \frac{\NF!}{\NC!(\NF-\NC)!}-1$.
The K\"ahler potential can now be expressed by
\beq
K_{U(1) \times SU(\NC)} =
\frac{\xi}{\NC} \log \det \left( QQ^\dagger \right)
= \frac{\xi}{\NC} \log \left( \sum_{\left<A\right>} \left|B^{\left<A\right>}\right|^2\right).
\label{eq:Kahler_U(N)_N=1}
\eeq
The $1/\NC$ factor in front is the (inverse) $U(1)$ charge of the
invariant $B^{\left<A\right>}$.
The FI parameter plays an important role: it forces the gauge
symmetry $U(1) \times SU(\NC)$ to be fully broken, namely it hides the
singularity at the origin, where the gauge symmetry is recovered. 

The Grassmannian manifold is one of the Hermitian symmetric spaces.
NL$\sigma$Ms on all Hermitian symmetric spaces 
can be obtained by imposing proper holomorphic constraints from $F$-terms,  
by which Hermitian symmetric spaces are embedded 
into ${\mathbb C}P^{\NF-1}$ or the Grassmannian manifold
\cite{Higashijima:1999ki}.

\subsection{The $U(\NC)$ Hyper-K\"ahler Quotient}

One can easily extend the above K\"ahler quotient to the hyper-K\"ahler
quotient by considering a natural ${\cal N}=2$ supersymmetric
extension. Here we study the $U(1) \times SU(\NC)$ case. 
The K\"ahler potential and the superpotential are given by 
\beq
\tilde K_{U(1) \times SU(\NC)} &=& \Tr \left[
Q Q^\dagger e^{-V_e}e^{-V'} + \tilde Q^\dagger \tilde Q e^{V_e}e^{V'} 
\right] + \xi V_e\ ,\label{eq:kahler_U(N)}\\
W &=& \Tr \left[Q\tilde Q  \Sigma \right]\ ,
\eeq
respectively, 
where we have introduced $\NF$ hypermultiplets $(Q,\tilde Q^\dag)$ in the
fundamental representation of $U(\NC) \simeq U(1) \times SU(\NC)$ and
$U(\NC)$ vector superfields $(V,\Sigma) = (V'+V_e{\bf 1}_\NC,\Sigma)$.
The complexified gauge transformation is given by
\beq
Q \to e^{i\Lambda} Q\ ,\quad
\tilde Q \to \tilde Q e^{- i\Lambda}\ ,\quad
e^{V} \to e^{i\Lambda} e^V e^{- i\Lambda^\dagger}\ ,\quad
\Sigma \to e^{i\Lambda} \Sigma e^{- i\Lambda}\ ,\qquad
\Lambda \in GL(\NC,{\mathbb C})\ .
\eeq
The target space of the corresponding NL$\sigma$M is 
a hyper-K\"ahler manifold, namely the cotangent bundle 
$T^* Gr_{\NF,\NC}$ over the complex Grassmannian manifold $Gr_{\NF,\NC}$, 
endowed with the Lindstr\"om-Ro\v{c}ek metric \cite{Lindstrom:1983rt}. 
Let us obtain the K\"ahler potential with respect to $Q,\tilde Q$
without choosing the Wess-Zumino gauge.
The equations of motion for $\Sigma$ and $V$ are
\beq
Q \tilde Q = 0\ ,\label{eq:eom_U(N)_1}\\
-QQ^\dagger e^{-V} + e^V \tilde Q^\dagger\tilde Q + \frac{\xi}{\NC}
{\bf 1}_\NC = 0\ .\label{eq:eom_U(N)_2}
\eeq
The first equation implies that $\tilde Q$ is orthogonal to $Q$.
The rank of $Q$ must be $\NC$ due to the positive FI parameter $\xi$,
while $\tilde Q$ can be zero.
Therefore $Q\ (\tilde Q=0)$ parametrizes the base space $Gr_{\NF,\NC}$
with the total space being the cotangent bundle over it.  
Let us count the complex dimensions of the target space: $\NC\NF + \NF \NC
- \NC^2 - \NC^2 = 2\NC(\NF-\NC)$ where the first subtraction is the
$U(\NC)^{\mathbb C}$ quotient and the second is the number of conditions
given in Eq.~(\ref{eq:eom_U(N)_1}). 
In order to solve the second matrix equation, we first multiply by 
$\sqrt{QQ^\dagger}e^{-V}$ from the left and by
$\sqrt{QQ^\dagger}$ from the right\footnote{
Note that 
the square root and the logarithm is uniquely defined for 
positive (semi-)definite Hermitian matrices.  
This point might be missed (at least in this context) 
in the physics literature so far.
}, 
such that the matrix equation becomes Hermitian  
\beq
X^2 - \frac{\xi}{\NC} X - \sqrt{QQ^\dagger} \tilde Q^\dagger\tilde
Q\sqrt{QQ^\dagger} = 0\ , 
\qquad X \equiv \sqrt{QQ^\dagger}e^{-V}\sqrt{QQ^\dagger}\ .
\eeq
Therefore, using $\det QQ^\dagger\not=0$, we find the solution 
\beq
V&=& -\log \left[ 
\frac{1}{\sqrt{QQ^\dagger}}X
\frac{1}{\sqrt{QQ^\dagger}}
\right]\ ,\nn
\quad {\rm with~}\ 
X &=& \frac{\xi}{2\NC}{\bf 1}_\NC 
+ \sqrt{\sqrt{QQ^\dagger} \tilde Q^\dagger\tilde Q\sqrt{QQ^\dagger} +
  \frac{\xi^2}{4\NC^2}{\bf 1}_\NC}\ . 
\label{eq:sol_VU}
\eeq

We will now switch to another description i.e.~using holomorphic
invariants. 
We have the following invariants of the $SU(\NC)^{\mathbb C}$ gauge
group 
\beq
B^{\left<A\right>} = \det Q^{\left<A\right>}\ ,\quad
M = \tilde Q Q\ ,\quad 
\left(\tilde B_{\left<A\right>} = \det \tilde
Q_{\left<A\right>}\right)\ .
\label{eq:invariant_U}
\eeq
In addition to the Pl\"ucker relations for the $B^{\left<A\right>}$'s, 
there are constraints on the mesonic invariant $M$ 
\begin{eqnarray}
&& \quad M_B{}^{[A_1}B^{\langle A_2\cdots A_{\NC+1}]\rangle}=0\ ,\quad
B^{\langle A_1\cdots A_{\NC-1}A'\rangle}M_{A'}{}^B=0\ .
\end{eqnarray} 
Furthermore, $B^{\left<A\right>}$ (and $\tilde B_{\left<A\right>}$) 
are only defined up to $U(1)^{\mathbb C}$ equivalence
transformations. 
After reconstructing $Q$ from (some) non-vanishing
$B^{\left<A\right>}$, we can reconstruct $\tilde Q$ from the first
condition and find the constraint $Q\tilde Q=0$ from the second. 
Therefore, these invariants and their constraints describe 
the same target space, $T^* Gr_{\NF,\NC}$.
Plugging back the solution (\ref{eq:sol_VU}) into the K\"ahler
potential (\ref{eq:kahler_U(N)}), we obtain the K\"ahler potential in
terms of these invariants \cite{Lindstrom:1983rt,Antoniadis:1996ra}
\begin{align}
 \tilde K_{U(1) \times SU(\NC)} =&\ 
K_{U(1) \times SU(\NC)} \label{eq:HK_U(N)} \\
&+\frac{\xi}{\NC}\, {\rm Tr}_{\rm F}
\left[\sqrt{{\bf 1}_{\NF}+\frac{4\NC^2}{\xi^2}M M^\dagger }
-\log \left({\bf 1}_{\NF} + 
\sqrt{{\bf 1}_{\NF}+\frac{4\NC^2}{\xi^2}M M^\dagger }\right)\right]\ .
\nonumber
\end{align}
We have used $A^\dagger A=MM^\dagger$ and the cyclic property of a trace, i.e.~for
$A = \sqrt{QQ^\dagger} \tilde Q^\dagger$  
\begin{eqnarray}
 {\rm Tr}\left[f(AA^\dagger)- f({\bf 0}_{\NC}){\bf 1}_\NC\right] = {\rm Tr}\left[f(A^\dagger A)- f({\bf 0}_{\NF}){\bf 1}_{\NF}\right]\ .
\end{eqnarray}
This relation can be easily proved by expanding the function $f$ around $AA^\dagger= \mathbf 0_\NC$.
Recall that the logarithm and the square root of a positive
(semi$\textrm{-})$definite
Hermitian matrix can be calculated by diagonalization and 
therefore the cyclic property works not only for polynomial functions
but for any function $f(x)$.  

The hyper-K\"ahler quotient construction of 
the cotangent bundle over the Grassmann manifold has been reviewed here.
For $\NC=1$, the $U(1)$ hyper-K\"ahler quotient reduces to the 
cotangent bundle over the complex projective space, 
$T^*{\mathbb C}P^{\NF-1}$ \cite{Curtright:1979yz}, 
endowed with the Calabi metric \cite{Ca}. 
The explicit K\"ahler potentials of the cotangent bundles over 
the other Hermitian symmetric spaces have recently been obtained 
by a rather different method \cite{Gates:1998si}. 
It is an open question if these manifolds 
can be obtained as a certain hyper-K\"ahler quotient or not.

We will not repeat the derivation of the $SU(\NC)$ hyper-K\"ahler
quotient here. Explicit expressions can be found in the literature,
see for instance \cite{Antoniadis:1996ra,Arai:2003tc}.
It gives the cotangent bundle over the $SU(\NC)$ K\"ahler
quotient derived in the last subsection.

\section{The $SO(\NC)$ and $USp(2\MC)$ (Hyper-)K\"ahler Quotients}
\label{sec:kahlerquotient}

\subsection{The $SO(\NC)$ and $USp(2\MC)$ K\"ahler Quotients
  \label{sec:kahler_so_usp}} 

The K\"ahler potential for an $SO(\NC)$ or a $USp(2\MC)$ gauge theory 
is given by
\beq
 K_{SO,USp} = \Tr\left[QQ^\dagger e^{-V'}\right] \ , 
 \label{eq:Kahler_SOUSp}
\eeq
where $V'$ takes a value in the $\mathfrak{so}(\NC)$ or
$\mathfrak{usp}(2\MC)$ algebra. 
The $D$-flatness conditions in the Wess-Zumino gauge are 
\beq
 D^A = \Tr_{\rm F} \left(Q_{\rm wz}^\dagger T^A Q_{\rm wz}\right) = 0\ , 
  \label{eq:D-flat-SOUSp}
\eeq
with $T_A$ being the generators in the Lie algebra of $SO$ or $USp$. 

Instead of solving these equations explicitly, we will here discuss
the breaking pattern of the gauge symmetry and the flat directions.
For this we will use both the gauge and the global symmetry 
as is usually done.
The vacuum expectation value of $Q_{\rm wz}^{SO}$ in the case of 
$SO(\NC)$ can be put on the diagonal form after fixing both the local
and the global symmetries as \cite{Intriligator:1995id} 
\begin{align}
Q_{\rm wz}^{SO(\NC)} =
\left(A_{\NC \times \NC},\ {\bf 0}_{\NC \times (\NF-\NC)}\right)\ , \quad
     {\rm with} \quad A_{\NC \times \NC}
= {\rm diag}(a_1,a_2,\cdots,a_{\NC}) \ , 
\label{eq:SOvacuum}
\end{align}
where we have taken a normal basis for the $SO(\NC)$ group, namely
$g^Tg = {\bf 1}_{\NC}$. 
Here all the parameters $a_i$ are taken to be real and positive, which
indeed parametrize the flat directions of the Higgs branch. 
In generic points of the moduli space of vacua with non-degenerate $a_i$,
the gauge symmetry is completely broken 
and the flavor symmetry $U(\NF)$ is broken to $U(\NF-\NC)$. 
The moduli space of vacua 
can be locally written in generic points as
\begin{eqnarray}
 {\cal M}_{SO(\NC)} \simeq {\mathbb R}^{\NC}_{>0} \times 
\frac{U(\NF)}{U(\NF-\NC)
\times ({\mathbb Z}_2)^{\NC-1}
}\ . 
 \label{eq:SO-KQ}
\end{eqnarray}
Here the discrete unbroken group 
$({\mathbb Z}_2)^{\NC-1}$ has elements 
of $\NC$-by-$\NC$ diagonal matrices in the $SO(\NC)$ group elements 
acting from the left, 
which have an even number of $-1$ elements with the rest $1$, 
in addition to the same matrices embedded into the $U(\NF)$ group
acting from the right. 
We see that the space is of cohomogeneity $\NC$, 
of which the isometry is $U(\NF)$ and the isotropy at generic points 
is $U(\NF-\NC)$. 
The coordinates of the coset space $U(\NF)/U(\NF-\NC)$ correspond to 
Nambu-Goldstone (NG) modes of the broken flavor symmetry, 
whereas the coordinates $\{a_i\}$ of the flat directions 
${\mathbb R}_{>0}^{\NC}$ 
correspond to the so-called ``quasi-Nambu-Goldstone'' modes
\cite{Bando:1983ab}. 
The quasi-NG modes do not correspond to a symmetry breaking 
but are ensured by supersymmetry.
In general, the unbroken flavor symmetry, namely the isotropy of the space, 
changes from point to point depending on the values of 
the parameters (the quasi-NG modes) $a_i$'s.
When two parameters coincide, $a_i=a_j, (i\not =j)$, 
a color-flavor locking $SO(2)$ symmetry emerges.  
In such degenerate subspace on the manifold, 
the above coset space attached to ${\mathbb R}_{>0}^{\NC}$ shrinks to one
with less dimension;
${\cal M}_{SO(\NC)} \sim {\mathbb R}^{\NC+1}_{>0} \ltimes 
\frac{U(\NF)}{U(\NF-\NC)\times SO(2)\times (\mathbb{Z}_{2})^{\NC-2}}$.\footnote{
Some quasi-NG modes change to NG modes reflecting further symmetry breaking.
This change of quasi-NG and NG modes was pointed out in
Ref.~\cite{Kotcheff:1988ji}.
It was also observed in the moduli space of domain walls \cite{Eto:2008dm}
and of non-Abelian vortices \cite{Eto:2004ii}, 
where quasi-NG modes correspond to the positions of solitons.
Here the notation ``$\ltimes$" is used for a local structure of the bundle 
$F \ltimes B$ with a fiber $F$ and a base space $B$. 
This is not globally true; once some values of ${\mathbb R}_{>0}^\#$ change, the coset space changes in general. 
}
In general, when $n_i$ ($i=1,2,\cdots$, and 
$\sum_i n_i \leq \NC$) parameters among $a_i$ coincide, 
the symmetry structure of the moduli space of vacua becomes 
\begin{eqnarray}
 {\cal M}_{SO(\NC)} \sim 
{\mathbb R}^{\NC + \sum_i {\frac{1}{2}} n_i (n_i-1)}_{>0} \ltimes 
\frac{U(\NF)}{U(\NF-\NC) \times \prod_i SO(n_i)
\times ({\mathbb Z}_2)^{\NC-1-\sum (n_i-1)}}\ .
 \label{eq:SO-KQ2}
\end{eqnarray}
The most symmetric vacuum, when all parameters coincide, is realized as
\begin{eqnarray}
 {\cal M}_{SO(\NC)} \sim 
{\mathbb R}^{\frac{1}{2} \NC (\NC+1)}_{>0} \ltimes 
\frac{U(\NF)}{U(\NF-\NC) \times SO(\NC)}\ .
\end{eqnarray}
This breaking pattern of the flavor symmetry 
is the one of non-supersymmetric $SO(\NC)$ QCD \cite{Benson:1994dp}. 
The unbroken flavor symmetry in non-supersymmetric QCD 
is in general further broken down as in Eq.~(\ref{eq:SO-KQ}) or
(\ref{eq:SO-KQ2}) in supersymmetric QCD.

No singularities appear in the moduli space even when the parameters coincide 
unless they vanish.
The existence of the quasi-NG modes is strongly related to the
emergence of the Coulomb phase. 
When one $a_i$ vanishes, the NG part becomes 
$U(\NF)/U(\NF-\NC+1)$ but the gauge symmetry is still completely broken. 
Accordingly, no singularities appear.
However, when any two of the $a_i$'s vanish, 
an $SO(2)$ subgroup of the gauge symmetry is recovered
and the NG part becomes $U(\NF)/U(\NF-\NC+2)$.
(One expects a singularity on the manifold in the limit of two
vanishing $a_i$'s). 
Thus, in the Higgs phase with completely broken gauge symmetry, 
the rank of $Q_{\rm wz}$ has to be greater than $\NC-2$. 
In this paper we consider this latter case, the models with $\NF \ge
\NC-1$.

For the $USp(2\MC)$ case  
it is known that the flat directions are parametrized by \cite{Intriligator:1995ne,Argyres:1996hc}
\beq 
Q_{\rm wz}^{USp(2\MC)} = {\bf 1}_2 \otimes \left(A_{\MC \times \MC},\ {\bf 0}_{\MC \times (\MF-\MC)}\right)  \ , 
\label{eq:USpvacuum}
\eeq
where the number of flavors is even $\NF = 2\MF$.
Even in generic points with non-degenerate $\{a_i\}$,
color-flavor symmetries $USp(2)^{\MC} \simeq SU(2)^{\MC}$ exist in the vacuum.
Therefore, the moduli space of vacua can be locally written in generic points 
as 
\begin{eqnarray}
 {\cal M}_{USp(2\MC)} \simeq {\mathbb R}_{>0}^{\MC} \times 
 \frac{U(\NF)}{U(\NF-2\MC) \times USp(2)^{\MC} }\ ,
 \label{eq:USp-KQ}
\end{eqnarray}
except for submanifolds where the coset space shrinks.
The resulting space is of cohomogeneity $\MC$.
Again, when $n_i$ ($i=1,2,\cdots$, and 
$\sum_i n_i \leq \MC$) parameters among $a_i$ coincide, 
the symmetry structure becomes 
\begin{eqnarray}
 {\cal M}_{USp(2\MC)} ~\sim~
{\mathbb R}^{\MC + 2 \sum_i  n_i (n_i-1)}_{>0} \ltimes 
\frac{U(\NF)}{U(\NF-2\MC) \times USp(2)^{\MC - \sum_i n_i} \times 
  \prod_i USp(2n_i)}\ .
\end{eqnarray}
The most symmetric vacuum, when all parameters coincide, is realized as
\begin{eqnarray}
 {\cal M}_{USp(2\MC)} ~\sim~
{\mathbb R}^{\MC (2\MC-1)}_{>0} \ltimes 
\frac{U(\NF)}{U(\NF-2\MC) \times USp(2\MC)}\ ,
\end{eqnarray}
whose breaking pattern is the one of non-supersymmetric $USp(2\MC)$ QCD.
There are no singularities unless one of the parameters $a_i$
vanishes.
In the case of $USp(2\MC)$ the complete broken gauge symmetry needs
$\MF \ge \MC$.

Next we explicitly construct the K\"ahler potentials from
the moduli space of vacua.
The $D$-flatness conditions (\ref{eq:D-flat-SOUSp}), however,   
are rather difficult to solve.\footnote{
To our knowledge the $D$-flatness conditions are 
not solved in the case of an $SO$ or a $USp$, ${\cal N}=1$
supersymmetric gauge theory. }
Without taking the Wess-Zumino gauge, we can eliminate the superfield
$V'$ directly within the superfield formalism by using a trick. 
To this end we note that $V'$ satisfies $\det(e^{-V'})=1$ and
\beq 
V'{}^{\rm T}J+JV'=0\quad  \leftrightarrow
\quad e^{-V'{}^{\rm T}} J e^{-V'} = J\ . \label{eq:V'}
\eeq 
Here the matrix $J$ is the invariant tensor of the
$SO$ or $USp$ group, $g^{\rm T}J g=J$ with $g \in SO(\NC),\ USp(2\MC)$,
satisfying 
\beq
J^{\rm T} = \epsilon J\ ,\quad
J^\dagger J =  {\bf 1}_{\NC}\ ,\qquad
\epsilon = 
\left\{
\begin{array}{cl}
+1 & \text{for}\quad SO(\NC)\ ,\\
-1 & \text{for}\quad USp(\NC=2\MC)\ .
\end{array}
\right. \label{eq:def_J}
\eeq
We can choose the form of the invariant tensor $J$ as\footnote{
Two arbitrary choices of the invariant tensor 
are related by an appropriate unitary transformation $u$ :
$J'=u^{\rm T}J\,u$. Correspondingly, the elements of the gauge group 
for different choices of the invariant tensor are related by $g' = u^\dagger g u$. See Appendix \ref{sec:InvariantTensor}.}
\beq
J^{\pm}_{\MC}\equiv  
\left(
\begin{array}{cc}
{\bf 0}_{\MC} & {\bf 1}_{\MC}\\
\pm {\bf 1}_{\MC} & {\bf 0}_{\MC}
\end{array}
\right)\ , 
\qquad J_{\MC,{\rm odd}} \equiv \left(
\begin{array}{cc}
J_{\MC}^+ & {\vec 0}^{\,\rm T}\\
{\vec 0} & 1 
\end{array}\right) \ ,
\label{eq:repJ}
\eeq
where the last tensor is for the $SO(\NC=2\MC+1)$ case.
We will use these conventions throughout the paper unless otherwise stated. 

We are now ready to eliminate $V'$ using the following trick.
Let us first consider $V'$ taking a value in a larger algebra, namely
$\mathfrak{u}(\NC)$ and then introduce an $\NC$-by-$\NC$ matrix of Lagrange
multipliers\footnote{
Hermiticity of $\lambda$ is defined 
so that $\lambda e^{-V'{}^{\rm T}}J$ is a vector
superfield, that is, $\lambda^\dagger=e^{V'{}^{\rm T}}J \lambda\,
e^{-V'{}^{\rm T}}J$.}
$\lambda$
 to restrict $V'$ to take a value in the
$\mathfrak{so}(\NC)$ or the $\mathfrak{usp}(\NC=2\MC)$ subalgebra: 
\beq
 K_{SO,USp} = \Tr\left[ QQ^\dagger e^{-V'} + 
 \lambda \left(e^{-V'{}^{\rm T}} J e^{-V'} - J \right) \right]\ ,
 \label{eq:so_usp_kahler}
\eeq
where $Q$ are $\NF$ chiral superfields as earlier and $V'$ is a vector
superfield of $U(\NC)$. 
The added term breaks the complexified gauge transformation to 
$SO(\NC), USp(2\MC)$
 and the equation of motion for $\lambda$ gives the constraint (\ref{eq:V'})
which reduces the K\"ahler potential (\ref{eq:so_usp_kahler}) back to 
(\ref{eq:Kahler_SOUSp}). 
Instead, we will take another path and eliminate $V'$. 
The equation of motion for $V'$ takes the form
\beq
QQ^\dagger e^{-V'} + \left(\lambda + \epsilon\lambda^{\rm T}\right)J =
0\ , \label{eq:EOM_V'}
\eeq
where we have used (\ref{eq:V'}).
Combining (\ref{eq:EOM_V'}) with its transpose: $e^{-V'{}^{\rm
 T}}Q^*Q^{\rm T} + J(\lambda + \epsilon\lambda^{\rm T}) = 0$,
then  $\lambda$ can be eliminated:   
\beq
 QQ^\dagger e^{-V'} = e^{V'} J^\dagger Q^*Q^{\rm T}
 J\ . \label{SO,USp_eom_senza_lambda}
\eeq
Furthermore, in order to make the equation Hermitian, we multiply by
$\sqrt{QQ^\dagger}e^{-V'}$ from the left and by
$\sqrt{QQ^\dagger}$ from the right as in the previous
case 
\beq
X^2 =  \left(Q^{\rm T} J \sqrt{QQ^\dagger}\right)^\dagger \left(Q^{\rm
  T} J \sqrt{QQ^\dagger}\right)\ ,
\qquad
X \equiv \sqrt{QQ^\dagger}e^{-V'}\sqrt{QQ^\dagger}\ . \label{SO,USp_Xeq}
\eeq
This equation uniquely gives a positive definite matrix $X$, by
means of its square root. We can thus uniquely obtain $V'$ from this
$X$, if and only if the holomorphic invariants $M \equiv Q^T J Q$
satisfy ${\rm rank}\, M>\NC-2$, that is, if and only if the vacuum is
in the full Higgs phase. 
See Appendix \ref{app:unique_proof} for a uniqueness proof, in the
case of ${\rm rank}\,M=\NC-1$.
It is possible to switch to $Q_{\rm wz}$ from $Q$ by the complexified
gauge transformation $Q_{\rm wz} = u'{}^{-1} Q$ with $u'u'{}^\dagger =
 e^{V'}$.
Without using an explicit solution for $V'$, we obtain the K\"ahler
potential of the NL$\sigma$M
\beq
K_{SO,USp} = \Tr X =
\Tr \sqrt{\left(Q^{\rm T} J \sqrt{QQ^\dagger}\right)^\dagger
  \left(Q^{\rm T} J \sqrt{QQ^\dagger}\right)}\ .
\label{eq:kahlerpot_so_sp_phi}
\eeq
Thus we have obtained the explicit K\"ahler potentials. 

Now we can naturally switch to another expression for this NL$\sigma$M 
in terms of the holomorphic gauge invariants.
With the help of $\Tr\sqrt{AA^\dagger} = \Tr_{\rm F}\sqrt{A^\dagger
  A}$, one can rewrite the K\"ahler potential
(\ref{eq:kahlerpot_so_sp_phi}) as
\beq
K_{SO,USp}
= \Tr_{\rm F}\sqrt{M M^\dagger}\ ,\qquad
M^{\rm T} = \epsilon M\ ,
\label{eq:Kahler_SO,USp_N=1}
\eeq
where $M$ is nothing but the holomorphic invariants of the gauge
symmetry
\beq
 M \equiv Q^{\rm T} J Q\ ,\qquad
 B^{\left<A\right>} \equiv \det Q^{\left<A\right>}\ .
  \label{eq:meson-baryon}
\eeq
The first one is the ``mesonic" invariant while the second is the
``baryonic" one which appears for $\NF \ge \NC$.
The two kinds of invariants should be subject to constraints in order
to correctly describe the NL$\sigma$M. 
There are relations between the mesons and the baryons:
\beq
&&SO(\NC):\det(J) \ B^{\left<A\right>} B^{\left<B\right>} =  
{\rm det} M^{\left<A\right>\left<B\right>},
\label{eq:rel_MB_SO}\\
&&USp(2\MC):{\rm Pf}(J)\ B^{\left<A\right>} = 
{\rm Pf}\,M^{\left<A\right>\left<A\right>}.
\eeq
where the $\NC$-by-$\NC$ matrix 
$M^{\left<A\right>\left<B\right>}$ is a minor matrix 
defined by $\left(M^{\left<A\right>\left<B\right>}\right)^{ij}=M^{A_iB_j}$.
The Pl\"ucker relations among the baryonic invariants
$B^{\left<A\right>}$ are derived from the above relation.
Actually, from the invariants $M$ and $B^{\left<A\right>}$ with the
constraints 
we can reconstruct $Q$ modulo the complexified gauge symmetry as follows.
By using an algorithm similar to the Cholesky decomposition of an
Hermitian matrix, we can show that
\begin{eqnarray}
&&\hbox{An arbitrary $n$-by-$n$ (anti-)symmetric complex matrix $X$ can }\nn 
&&\hbox{always be decomposed as $X=p^{\rm T}Jp$ with 
a ${\rm rank}(X)$-by-$n$ matrix $p$.}
\end{eqnarray}
See Appendix \ref{sec:DecompMatrix} for a proof of this statement.
In the $USp$ case, with a decomposition of the meson $M$,
we can completely reconstruct $Q$ modulo $USp(2\MC)^{\mathbb C}$
transformations.  
This fact corresponds to the fact that there are no independent 
baryons $B^{\left<A\right>}$ in this $USp(2\MC)$ theory and 
only the meson fields describe the full Higgs phase
\begin{eqnarray}
{\cal M}_{USp}=\left\{M\,|\,M\in \NF \mbox{-by-}\NF {~\rm matrix},\quad   
M^{\rm T}=-M,\quad {\rm rank}\,M=2\MC \right\}\ .
\end{eqnarray}
On the contrary, in the $SO(\NC)$ case, 
a decomposition of $M$ gives $Q$ modulo $O(\NC)^{\mathbb C}$
and one finds two candidates for $Q$ since 
${\mathbb Z}_2\simeq O^{\mathbb C}/SO^{\mathbb C}$ which is
fixed by the sign of the baryons.\footnote{
In the case of ${\rm rank}\,M=\NC-1$, 
$g\in {\mathbb Z}_2$ acts trivially on
$Q$ as $g\,Q=Q$, although all the baryons vanish.} 
Therefore we have to take the degrees of freedom of the baryons into
account to consider the full Higgs phase
\begin{align}
 {\cal M}_{SO}=
\left\{M, B^{\left<A\right>}\,|\,M:\,{\rm symmetric}~\NF \mbox{-by-} \NF,\, 
{\rm Eq.~(\ref{eq:rel_MB_SO})},\, \NC-1\le {\rm rank}\,M\le
\NC\right\}\ .
\quad
\end{align}

For large $\NC$, it is a hard task to obtain an 
explicit metric from the formula (\ref{eq:Kahler_SO,USp_N=1}), since
we need to calculate the eigenvalues of $MM^\dagger$. 
Let us, therefore, consider expanding the K\"ahler potential 
(\ref{eq:Kahler_SO,USp_N=1})
in terms of infinitesimal coordinates around a point.
Note that the meson field $M$ for $SO(\NC)$, 
which is a symmetric matrix, can always be diagonalized 
by using the flavor symmetry $U(\NF)$ as  
\begin{eqnarray}
M_{\rm vev}^{SO}\equiv u M u^{\rm T}
={\rm diag}(\mu_1,\mu_2,\cdots,\mu_{\NC},0,\cdots)\ ,  
\end{eqnarray}
with $u\in U(\NF)$ and  parameters 
$\mu_i\in {\mathbb R}_{\ge 0}$ are square roots of the eigenvalues of
$MM^\dagger$.
The meson field $M$ in the $USp(2\MC)$ case, which is an anti-symmetric
matrix, can be also diagonalized as
\begin{eqnarray}
M_{\rm vev}^{USp}\equiv u M u^{\rm T}
= \left(
\begin{array}{cc}
 0&1 \\-1&0
\end{array}\right) \otimes
{\rm diag}(\mu_1,\mu_2,\cdots,\mu_{\MC},0,\cdots)\ .
\end{eqnarray}
See Appendix \ref{sec:DecompMatrix} for the proof.
These vacuum configurations in both the cases,
$M_{\rm vev}=M_{\rm vev}^{SO},\,M_{\rm vev}^{USp}$, 
are summarized as 
\begin{eqnarray}
 (M_{\rm vev})_{ij}=\mu_i (J)_{ij}=(J)_{ij}\mu_j \ ,
\end{eqnarray}
where we take the invariant tensors 
as $(J)_{ij}=\delta_{ij}$ for the $SO(\NC)$ case, and 
$(J)_{ij}=\delta_{i+\MF,j}-\delta_{i,j+\MF}$ 
and $\mu_{i+\MF}\equiv \mu_i, (1\le i\le \MF)$ 
in the case of $USp(\NC=2\MC)$. 

For simplicity, let us concentrate on
the $SO(\NC)$ case with $\NC=\NF$, and consider generic points
of the manifold with ${\rm rank}(M_{\rm vev})=\NC$, 
that is, $\mu_i>0$ for all $i$.  
In this case, there are no constraints for the meson field locally,
and thus, the meson field $M$ can be treated as 
coordinates  parametrizing the manifold locally.
  It is convenient to consider 
a small fluctuation $\phi=M-M_{\rm vev}$ around the vacua $M_{\rm vev}$
and expand the formula (\ref{eq:Kahler_SO,USp_N=1}) 
with respect to $\phi$.
The following formula is useful to expand a function  $f(X)$ 
of a matrix $X$ in a trace around
$X=X_0$, 
\begin{eqnarray}
 {\rm Tr}[f(X_0+\delta X)] &=& \frac{1}{2\pi i} \oint_{\cal C} d\lambda \ 
f(\lambda) {\rm Tr} \left[\frac{\bf 1}{\lambda {\bf 1}-X_0-\delta X}\right]\nn
&=&{\rm Tr}[f(X_0)]+\sum_{n=1}^\infty\frac{1}{2\pi n\, i}
\oint_{\cal C}d\lambda \ 
f'(\lambda){\rm Tr}\left[\left(\frac{\bf 1}{\lambda {\bf 1}-X_0}\delta X \right)^n\right],\label{eq:exp_formula}
\end{eqnarray}
where the closed path ${\cal C}$ surrounds 
all eigenvalues of $f(X)$ on the real positive axis
but no singularities of $f(\lambda)$. We set $f(\lambda)=\sqrt{\lambda}$ and
\begin{eqnarray}
X=MM^\dagger\ ,\quad 
X_0={\rm diag}(\mu_1^2,\cdots,\mu_{\NC}^2)\ ,\quad 
\delta X=M_{\rm vev}\phi^\dagger+\phi M^\dagger_{\rm
  vev}+\phi\phi^\dagger\ .
\end{eqnarray}
Since $f(\lambda) = \sqrt{\lambda}$ has a branch point at the origin, 
the eigenvalues $\mu_i$ cannot be zero in this formula.
To proceed the calculation, we need to perform the integrations
\begin{eqnarray}
A_n(\mu_1,\cdots,\mu_n)\equiv
\frac{1}{2\pi i}\oint \frac{d\lambda}{\sqrt{\lambda}}
\prod_{i=1}^n\frac{1}{\lambda-\mu_i^2}\ .
\end{eqnarray}
The results of the integrations can be expressed in terms of 
the elementary symmetric polynomials, $C^{(m)}_{k_1k_2\cdots k_n},(m\le n)$
defined by 
\begin{eqnarray}
\prod_{i=1}^n(t+\mu_{k_i})=\sum_{m=0}^n C_{k_1\cdots
k_n}^{(m)} t^{n-m}\ ,\quad
 P_{k_1k_2\cdots k_n}\equiv \prod_{m>n}(\mu_{k_m}+\mu_{k_n})\ ,
\end{eqnarray}
where we also use a symmetric polynomial $P_{k_1\cdots k_n}$. 
The first few integrations give
\begin{eqnarray}
A_1(\mu_1)&=&\frac{1}{\mu_1}\ ,\quad 
A_2(\mu_1,\mu_2)=-\frac{1}{\mu_1 \mu_2(\mu_1+\mu_2)}\ ,\nn
A_3(\mu_1,\mu_2,\mu_3)&=&
\frac{C^{(1)}_{123}}{C^{(3)}_{123}P_{123}}
=\frac{\mu_1+\mu_2+\mu_3}{\mu_1
  \mu_2\mu_3(\mu_1+\mu_2)(\mu_2+\mu_3)(\mu_3+\mu_1)}\ ,\nn  
A_4(\mu_1,\mu_2,\mu_3,\mu_4)&=&-
\frac{C^{(1)}_{1234}C^{(2)}_{1234}-C^{(3)}_{1234}}
{C^{(4)}_{1234} P_{1234}}\ . 
\end{eqnarray}
After this preparation, we obtain the first few terms of
the expansion of the K\"ahler potential as
\begin{eqnarray}
 K_{SO}&=&\frac 12 \sum_{i,j}
\frac{\phi_{ij}\phi_{ji}^\dagger}{\mu_i+\mu_j}
\nonumber\\
&&{}-\frac12\sum_{i,j,k}\frac{\mu_i\, \phi_{ij}\phi^\dagger_{jk} \phi_{ki}}
{(\mu_i+\mu_j)(\mu_j+\mu_k)(\mu_k+\mu_i)}+{\rm c.c.}\nonumber\\
&&{}+\frac12\sum_{i,j,k,l}\frac{\mu_j\mu_k C^{(1)}_{ijkl}}{P_{ijkl}}
\phi_{ij}\phi_{jk} \phi_{kl}\phi^\dagger_{li}+{\rm c.c.}\nonumber\\
&&{}+\frac12\sum_{i,j,k,l}\frac{\mu_j\mu_lC^{(1)}_{ijkl}}{P_{ijkl}}
\phi_{ij}\phi_{jk} \phi_{kl}^\dagger\phi^\dagger_{li}
-\frac14\sum_{i,j,k,l}\frac{C^{(3)}_{ijkl}}{P_{ijkl}}
\phi_{ij}\phi_{jk}^\dagger\phi_{kl}\phi_{li}^\dagger\nonumber\\
&&{}+\hbox{K\"ahler~trf.}+{\cal O}(\phi^5)\ . \label{ExpKahlerSO}
\end{eqnarray}
A coordinate singularity emerges 
in the limit $\mu_i\rightarrow 0$ 
since the expansion formula (\ref{eq:exp_formula}) is not applicable
for $\mu_i=0$. The above result gives enough information to calculate
the scalar curvature $R$ of the manifold at $M=M_{\rm vev}$ in the
$SO(\NC)$ case, with a K\"ahler metric $g_{I\bar J}$
\begin{eqnarray}
 R|_{\phi=0}
&=&-2 g^{I\bar J}\partial_I\partial_{\bar J}\log\det g\Big|_{\phi=0}\nonumber\\
&=&2\sum_{i>j}\left(
\frac1{\mu_i+\mu_j}+\sum_k\frac{\mu_k}{(\mu_k+\mu_i)(\mu_k+\mu_j)}\right)
\,>0 \ ,
\end{eqnarray}
where the indices $I,\bar J$ label the components as 
$\phi^I=\phi_{ij},(i\ge j)$.
This result shows that the coordinate singularity with  
${\rm rank}(M_{\rm vev})=\NC-1$ 
can be removed by taking appropriate coordinates 
and, on the other hand,  
the submanifold with ${\rm rank}(M_{\rm vev})<\NC-1$ is a
curvature singularity of the manifold. 
That is, the curvature singularity lies in the region corresponding to 
the Coulomb phase of the original gauge theory, as we expected. 

The expansion of the K\"ahler potential in the $USp(2\MC)$ case, we
obtain the result (\ref{ExpKahlerSO}) with the substitution
$\phi\to\phi J^\dag,\ \phi^\dag\to J\phi^\dag$ and the curvature obtained
using this expanded potential reads
\beq
 R|_{\phi=0}=4\sum_{i>j}^{\MC}\left(
\frac{1}{\mu_i+\mu_j}+\sum_k^{\MC}
\frac{4\mu_k}{(\mu_k+\mu_i)(\mu_k+\mu_j)}
\right) \,>0 \ .
\eeq
This result shows that the submanifold with ${\rm rank}(M_{\rm
  vev})< 2(\MC-1)$ is a curvature singularity of the manifold. This
expansion, however, does not reveal the singularity appearing at ${\rm
  rank}(M_{\rm vev})=2(\MC-1)$. To detect this singularity, we
consider a deformation of the K\"ahler potential
\beq K_{USp,{\rm deformed}} = \Tr\,\sqrt{MM^\dag + \varepsilon^2} \ ,
\label{eq:Kuspdeform} \eeq
and make a similar expansion (see Appendix \ref{appuspdeform}). 
Taking now only one eigenvalue, say $\mu_1\to 0$ we find a term in the
scalar curvature 
\beq \lim_{\mu_1\to 0}\left.R\right|_{\phi=0} \supset
\frac{2}{\varepsilon} \ , \eeq
which shows the presence of a singularity for one vanishing
eigenvalue, that is corresponding to an unbroken $USp(2)\simeq
SU(2)$ symmetry.

\subsection{The $U(1) \times SO(\NC)$ and $U(1) \times USp(2\MC)$ K\"ahler Quotients}

Next, we would like to consider a K\"ahler quotient with gauging an
overall $U(1)$ phase in addition to the $SO(\NC)$ or $USp(2\MC)$ gauge
symmetry. 
We turn on the FI $D$-term associated with the additional $U(1)$ gauge
group. The K\"ahler potential can be written as
\beq
K_{U(1) \times (SO,USp)} = \Tr\left[ QQ^\dagger e^{-V'} e^{-V_e} + 
\lambda \left(e^{-V'{}^{\rm T}} J e^{-V'} - J \right) \right] + \xi
V_e\ , 
\eeq
where $V_e$ is the vector multiplet of the additional $U(1)$ gauge field.
We have already solved the $SO(\NC)$ and $USp(2\MC)$ part in the
previous section, so the K\"ahler potential can be rewritten as
\beq
K_{U(1) \times (SO,USp)} = \Tr\left[ \sqrt{ M M^\dagger}
  \right]e^{-V_e} + \xi V_e\ .
\eeq
The equation of motion for $V_e$ can be solved by
$
V_e = \log\left[{\Tr\left(\sqrt{ M M^\dagger}\right)}/{\xi}\right].
$
Plugging this into the K\"ahler potential, we obtain
\beq
K_{U(1) \times (SO,USp)} = \xi \log \left[\Tr\left(\sqrt{ M
    M^\dagger}\right)\right]\ ,\qquad
 M \equiv Q^{\rm T} J Q\ .
\label{eq:Kahler_pot_UxSO}
\eeq
In the case of $\NC=\NF$, we can expand the K\"ahler potential 
around a point $M=M_{\rm vev}$ by using the same method as
in Sec.\ref{sec:kahler_so_usp},
\begin{eqnarray}
 K_{U(1) \times (SO,USp)} &=& 
\frac{\xi}{2\sum_{k=1}^{\NC} \mu_k}\left(\sum_{i,j}^{\NC}
\frac{\phi_{ij}(\phi_{ij})^\dagger}{\mu_i+\mu_j}-\frac1{2\sum_{k=1}^\NC\mu_k}
\left|\sum_{i=1}^\NC(J^\dagger\phi)_{ii}\right|^2\right)\nn
&&{} +\hbox{K\"ahler trf.}
+{\cal O}(\phi^3)\ .\label{eq:Kahlermet_U1SOUSp}
\end{eqnarray}
Here we can confirm that 
the mode $\phi\propto M_{\rm vev}$ corresponding to $U(1)^{\mathbb C}$
is not effective in this K\"ahler potential. 
Therefore, with the constraint $\Tr\,[\phi J^\dag] = 0$, we can write the
K\"ahler potential to fourth order as
\begin{align}
K_{U(1)\times (SO,USp)} = 
\frac{\xi}{\sum_{k=1}^{\NC}\mu_k}
\Bigg[&K_{SO,USp} - \frac{1}{8\sum_{l=1}^{\NC}\mu_l}
\left|\sum_{i,j}\frac{\phi_{ij}\phi_{ji}^\dag}{\mu_i+\mu_j}\right|^2\\
&-\frac{1}{16\sum_{l=1}^{\NC}\mu_l}
\left|\sum_{i,j}\frac{(\phi J^\dag)_{ij}(\phi
 J^\dag)_{ji}}{\mu_i+\mu_j}\right|^2\Bigg] 
+ \ {\textrm{K\"ahler trf.}} + \mathcal{O}(\phi^5) \ . \nonumber
\end{align}
from which we obtain the curvatures as
\begin{eqnarray}
\xi  R_{U(1) \times (SO,USp)} = R_{(SO,USp)} \displaystyle
\sum_{i=1}^{\NC} \mu_i + 2\hat N_\epsilon(\hat N_\epsilon+1) \ ,
\end{eqnarray}
where $\hat N_\epsilon$ is the complex dimension of the manifold
\begin{eqnarray}
 \hat N_\epsilon \equiv 
{\rm dim}_{ \mathbb C} \ {\cal M}_{U(1) \times (SO,USp)}^{\rm vacuum}
=\frac{\NC(\NC+\epsilon)}2-1\ , \hs{5} \epsilon =
\left\{ \begin{array}{cc} + 1 & \mbox{for} \hs{5} SO \ , \\ -1 &
  \ \mbox{for} \hs{5} USp \ . \end{array} \right. 
\end{eqnarray}

A typical property of these theories is the existence of 
curvature singularities of the K\"ahler manifold. 
Since the Coulomb phase attached to the Higgs phase in the
original gauge theory is strongly related to a singularity,
the curvature singularity with $0<{\rm rank}(M)<\NC-1$ 
still survives after the $U(1)$ gauging for the case of $\NC\ge 3$,
while gauging $U(1)$ in the $SU(\NC)$ case removes the singularity.

\subsection{Examples}

\subsubsection{The $SO(2)$ Quotient (SQED) and the $U(1) \times SO(2)$
  Quotient \label{sec:example}}

The first example is $SO(2)$ with $\NF = 1$.
We have a complexified gauge symmetry $SO(2)^{\mathbb C}$, so the
corresponding target space is
\beq
{\cal M}^{SO(2)}_{\NF=1} = {Q}/\!\!\sim\ ,\quad
Q \sim g' Q\ ,\quad g'\in SO(2)^{\mathbb C}\ ,
\eeq 
where $Q = (Q_+,\ Q_-)^{T}$. 
In general, matrices in $SO(2)^{\mathbb C}$ can be expressed as
\beq
g' = 
\left(
\begin{array}{cc}
v' & 0\\
0 & 1/v'
\end{array}
\right)\ , 
\quad v' \in {\mathbb C}^*\ .
\eeq
This simply shows the fact that $SO(2) \simeq U(1)$ under which $Q_+$ has
charge $+1$ while $Q_-$ has charge $-1$. This is nothing else than
supersymmetric QED. 
The target space apparently seems to be a weighted complex projective 
space which is not a Hausdorff space
\beq
{\cal M}^{SO(2)}_{\NF=1} = W{\mathbb C}P^1_{(1,-1)}\ .
\label{eq:moduli_SO(2)_naive}
\eeq
However, we have to be careful. 
Sick points $(Q_+,Q_-) = (Q_+,0),\ (0,Q_-)$ for $Q_+\neq0$ and $Q_-\neq0$ are 
forbidden by the $D$-term condition $|Q_+|^2 - |Q_-|^2 = 0$ in the
Wess-Zumino gauge. 
To understand the true well-defined target space, 
we take the holomorphic invariant of this model to be
\beq
 M = 2Q_+Q_-\ .
\eeq
This is a good coordinate on the target space and the K\"ahler potential
is given by 
\beq
K^{SO(2)}_{\NF=1} = \left| M\right|\ .
\eeq
There is a conical singularity at the origin and the true target space
is 
\beq
{\cal M}^{SO(2)}_{\NF=1} = {\mathbb C}/{\mathbb Z}_2\ .
\eeq
At the singularity, the gauge 
symmetry is restored
and the vector multiplet obtains a massless field.
In general, singularities in a classical moduli space leads to the
appearance of some massless fields.
K\"ahler potentials usually acquire quantum corrections and they may
make such classical singular manifolds regular.

The second example is $U(1) \times SO(2)$ with $\NF = 1$. 
We turn on the FI parameters $\xi$ and we have
\beq
{\cal M}^{U(1) \times SO(2)}_{\NF=1} = {Q}/\!\!\sim\ ,\quad
Q \sim V_e V' Q\ ,\quad V_e \in U(1)^{\mathbb C}\ ,\quad V'\in
SO(2)^{\mathbb C}\ . 
\eeq
We can explicitly show that
\beq
g_eg' = 
\left(
\begin{array}{cc}
v_1 & 0\\
0 & v_2
\end{array}
\right),\quad v_1,v_2 \in {\mathbb C}^*\ .
\eeq
Here we impose that the gauge symmetry $U(1) \times SO(2)$ is free,
such that $|Q| \neq 0$. 
Hence, the target space is just one point.

Next, let us consider $\NF = 2$ with the $SO(2)$ and the $U(1) \times
SO(2)$ gauge groups. 
The scalar field is a 2 by 2 complex matrix
\beq
Q =
\left(
\begin{array}{cc}
Q_{+1} & Q_{+2}\\
Q_{-1} & Q_{-2}
\end{array}
\right) \equiv
\left(
\begin{array}{c}
\vec Q_+\\
\vec Q_-
\end{array}
\right)\ .
\eeq
The holomorphic invariants of the $SO(2)$ part are on the form
\begin{align}
 M_{SO(2)} &= \left\{Q^{\rm T} J Q,\ \det Q\right\} \nonumber\\
&=
\left\{
\left(
\begin{array}{cc}
2 Q_{-1}Q_{+1} & Q_{+1}Q_{-2} + Q_{+2}Q_{-1}\\
Q_{+1}Q_{-2} + Q_{+2}Q_{-1} & 2 Q_{+2}Q_{-2}
\end{array}
\right),\ 
Q_{+1}Q_{-2} - Q_{+2}Q_{-1}
\right\}.
\end{align}
We have to remove the points $\vec Q_+ = 0$ and $\vec Q_- = 0$,
where all the holomorphic invariants vanish $M = 0$.
The moduli spaces of vacua turn out to be
\beq
{\cal M}^{SO(2)}_{\NF=2} &=& W{\mathbb C}P^3_{(1,1,-1,-1)}-\{ M_{SO(2)} = 0\} = 
(({\mathbb C}^2)^*_+\times({\mathbb C}^2)^*_-)/{\mathbb C}^*\ ,\\
{\cal M}^{U(1) \times SO(2)}_{\NF=2} &=& 
\left(({\mathbb C}^2)^*/{\mathbb C}^*\right) \times \left(({\mathbb
  C}^2)^*/{\mathbb  C}^*\right)
= {\mathbb C}P^1 \times {\mathbb C}P^1. \label{eq:mod_vac_U(1)xSO(2)}
\eeq
Since positive real eigenvalues $\lambda_1$ and $\lambda_2$ 
satisfy
$\sqrt{\lambda_1}+\sqrt{\lambda_2}
=\sqrt{\lambda_1+\lambda_2+2\sqrt{\lambda_1\lambda_2}}$,
the K\"ahler potential can be easily shown to be
\beq
K^{SO(2)}_{\NF=2} &=& \sqrt{{\rm Tr}MM^\dagger+2\sqrt{\det MM^\dagger}}=
2 \sqrt{ |\vec Q_+|^2 |\vec Q_-|^2}\ ,\\
K^{U(1)\times SO(2)}_{\NF=2} &=& \frac{\xi}{2} \log |\vec Q_+|^2 +
\frac{\xi}{2} \log|\vec Q_-|^2\ .
\label{eq:kahler_SO(2)}
\eeq
The prefactor $\xi/2$ in Eq.~(\ref{eq:kahler_SO(2)}) will turn out
to have a significant difference from the usual prefactor $\xi$ of the
K\"ahler potential for usual ${\mathbb C}P^1$, see
Eq.~(\ref{eq:Kahler_U(N)_N=1}), 
when we will consider 1/2 BPS solitons.

It is straightforward to extend this to the case with generic $\NF$.
The manifolds are on the form
\beq
{\cal M}^{SO(2)}_{\NF} &=& W{\mathbb
  C}P^{2\NF-1}_{(1_{\NF},-1_{\NF})}-\{ M_{SO(2)} = 0\} =  
(({\mathbb C}^{\NF})^*_+\times({\mathbb C}^{\NF})^*_-)/{\mathbb C}^*\ ,\\
{\cal M}^{U(1)\times SO(2)}_{\NF} &=& 
\left(({\mathbb C}^{\NF})^*/{\mathbb C}^*\right) \times
\left(({\mathbb C}^{\NF})^*/{\mathbb C}^*\right) 
= {\mathbb C}P^{\NF-1} \times {\mathbb C}P^{\NF-1}\ .
\eeq
The K\"ahler potential for the latter manifold can be obtained by
merely replacing the two vectors $Q_{1,2}$ by $\NF$ vectors in 
Eq.~(\ref{eq:kahler_SO(2)}). Then the meson field becomes an
$\NF$-by-$\NF$ matrix, however, only two eigenvalues
$\lambda_1,\lambda_2$ of $MM^\dagger$ take non-zero values and in this
case we have the following identity
\beq
\det(\lambda{\bf 1}_{\NF}-MM^\dagger)=
\lambda^{\NF-2}
\det\left(\lambda {\bf 1}_2-(QQ^\dagger)J^\dagger (QQ^\dagger)^{\rm
  T}J\right) \ .
\eeq
From this characteristic polynomial, we can read off
\begin{eqnarray}
 \lambda_1+\lambda_2=2|\vec Q_+|^2|\vec Q_-|^2+2|\vec Q_+ \vec
 Q_-^\dagger|^2\ , \qquad
\lambda_1\lambda_2 =\left(|\vec Q_+|^2|\vec Q_-|^2-|\vec Q_+
\vec Q_-^\dagger|^2\right)^2\ .
\end{eqnarray}
Therefore, we find also in the case of $\NF$ flavors
\begin{eqnarray}
K_{\NF}^{SO(2)} = \sqrt{\lambda_1}+\sqrt{\lambda_2} = 
2\sqrt{|\vec{Q}_+|^2|\vec{Q}_-|^2}\ .
\end{eqnarray}

\subsubsection{The $USp(2)$ Quotient}
This case completely reduces to the $SU(2)$ case with $\NF$ flavors. 
It is not difficult to show that only two eigenvalues
$\lambda_1,\lambda_2$ of $MM^\dagger$ take non-zero values and they
coincide
\begin{eqnarray}
\lambda_1=\lambda_2=\frac12 {\rm Tr}[MM^\dagger]={\rm
  det}(QQ^\dagger)\ ,
\end{eqnarray}
and this indeed yields the K\"ahler potential for the $SU(2)$ case
\begin{eqnarray}
K_{\NF}^{USp(2)\simeq SU(2)} = {\rm
  Tr}[\sqrt{MM^\dagger}]=2\sqrt{\det(QQ^\dagger)}\ . 
\end{eqnarray}
We find explicitly the ${\mathbb Z}_2$-conifold singularity at the
origin in this model.

\subsubsection{The $USp(4)$ Quotient}

By ``diagonalizing'' $M$ by $M_{ij}=\mu_iJ_{ij}$, we find two
non-vanishing eigenvalues both with multiplicity two, that is
$\lambda_1=\lambda_3=\mu_1^2$ and $\lambda_2=\lambda_4=\mu_2^2$ and
they can be written as
\begin{eqnarray}
 \lambda_1+\lambda_2=\frac12 {\rm Tr}[MM^\dagger]\ ,
\quad \lambda_1\lambda_2=\sum_{\langle A\rangle}|P_{\langle
  A\rangle}|^2\ ,
\label{eq:eigenvaluesUsp4}
\end{eqnarray}
where $P_{\langle A\rangle}$ is the Pfaffian of a minor matrix
\begin{eqnarray}
 P_{\langle A_1A_2A_3A_4\rangle}\equiv 3 M_{A_1[A_2}M_{A_3A_4]}\ .
\end{eqnarray}
In this case where we have $USp(4)$ i.e.~$\MC=2$, thus it can be written as
\begin{eqnarray}
\sum_{\langle A\rangle}|P_{\langle A\rangle}|^2
=\frac18\left({\rm Tr}[MM^\dagger]\right)^2-\frac14 {\rm
  Tr}[(MM^\dagger)^2] \ .
\end{eqnarray}
Since the right hand sides of both the equations in
Eq.~(\ref{eq:eigenvaluesUsp4}) are invariant under the flavor 
transformation performing the diagonalization, we find for generic
number of flavors $\NF$
\begin{eqnarray}
K_{\NF}^{USp(4)} = 2\left(\sqrt{\lambda_1}+\sqrt{\lambda_2}\right)
= 2\sqrt{\frac12{\rm Tr}[MM^\dagger]
+ 2\sqrt{\sum_{\langle A\rangle}|P_{\langle A\rangle}|^2}}\ . 
\end{eqnarray}
Considering a minimal case with $\MF=\MC=2$, with the following
parametrization
\begin{eqnarray}
 M=\left(
\begin{array}{cccc}
 0& \phi_1& \phi_2& \phi_3\\
-\phi_1&0 &\chi_3&-\chi_2\\
-\phi_2&-\chi_3&0&\chi_1\\
-\phi_3&\chi_2&-\chi_1&0
\end{array}\right) \ , 
\end{eqnarray}
we find ${\rm Pf}(M)=\vec \phi\cdot\vec \chi$ and the simple form of
the K\"ahler potential
\begin{eqnarray}
K_{\NF=4}^{USp(4)} = 2\sqrt{\frac12{\rm Tr}[MM^\dagger]+2|{\rm Pf}(M)|}
=2\sqrt{|\vec\phi|^2+|\vec \chi|^2+2|\vec \phi\cdot\vec \chi|} \ .
\end{eqnarray}
Manifestly, we can observe an orbifold singularity on the submanifold
\begin{eqnarray}
|\vec\phi|^2+|\vec \chi|^2\not=0 \ ,
 \quad {\rm Pf}(M)=\vec \phi\cdot\vec \chi = 0 \ , 
\end{eqnarray}
of which the rank is $2\MC-2=2$, since the ${\rm Pf}(M)\in {\mathbb
  C}$ is an appropriate coordinate describing 
the orthogonal direction to the submanifold and 
the term $\sqrt{|{\rm Pf}(M)|^2}$ emerges in the potential.
In a generic region away from this singular submanifold, the scalar
curvature is given by 
\begin{eqnarray}
R = \frac{20}{\sqrt{|\vec\phi|^2+|\vec \chi|^2+2|\vec \phi\cdot\vec
    \chi|}} \ ,
\end{eqnarray} 
and is finite even in the vicinity of the submanifold.

\subsubsection{The $SO(3)$ Quotient}

The K\"ahler quotient for $SO(3)$ with $\NF$ flavors reads
\begin{eqnarray}
 K_{\NF}^{SO(3)} = \sqrt{\lambda_1}+\sqrt{\lambda_2}+
\sqrt{\lambda_3} \ ,
\end{eqnarray}
and it is obtained by solving the following algebraic equations
\begin{eqnarray}
 (K^2-A_1)^2=4 A_2+8 \sqrt{A_3} K \ ,
\end{eqnarray}
where the definitions are
\begin{eqnarray}
 A_1&\equiv&\lambda_1+\lambda_2+\lambda_3={\rm Tr}[MM^\dagger]\ ,\nn
A_2&\equiv&\lambda_1\lambda_2+\lambda_3\lambda_2+\lambda_3\lambda_1
=\frac12({\rm Tr}[MM^\dagger])^2-\frac12{\rm Tr}[(MM^\dagger)^2]\ ,\nn
A_3&\equiv&\lambda_1\lambda_2\lambda_3 \ .
\end{eqnarray}
A solution with a real number satisfying $K^2\ge A_1>0$ should be
unique. 
Here $\sqrt{A_3}$ does not imply a singularity immediately.
In the case of $\NF=\NC=3$, we can rewrite it in terms of the baryon
field $B$ as 
\begin{eqnarray}
 \sqrt{A_3}=\sqrt{\det(MM^\dagger)}=\sqrt{|\det M|^2 }=|B|^2 \ ,
\end{eqnarray}
and around the submanifold with $B=0$, 
$B$ is an appropriate coordinate around the submanifold.
With $K_0=K|_{|B|^2=0}$, we find
\begin{eqnarray}
 K_{\NF=3}^{SO(3)}=K_0+\frac{2|B|^2}{K_0^2-A_1}+{\cal O}(|B|^4)\ .
\end{eqnarray}
Since $K_0^2-A_1=0$ implies that $A_2=|B|^2=0$, which in turn implies
that ${\rm rank}\,M\ge \NC-2=1$, 
this expansion tells us that 
the submanifold with ${\rm rank}\,M=\NC-1=2$ is not singular.

Let us now consider this simple example of $SO(3)$ with $\NF = 2$. The
result of the K\"ahler potential is the same as in the $SO(2)$ case
with $\NF = 2$
\beq K_{\NF=2}^{SO(3)} = \sqrt{\Tr\,MM^\dag +2|\det M|} \ . \eeq

\subsection{The $SO(\NC)$ and $USp(2\MC)$ Hyper-K\"ahler Quotients}

Our next task is lifting up the $SO(\NC)$ and $USp(\NC = 2\MC)$
K\"ahler quotients of the previous subsection to the hyper-K\"ahler
quotients as we did for the $U(\NC)$ (hyper-)K\"ahler quotient in
Sec.~\ref{sec:review}. We leave the issues of the hyper-K\"ahler 
quotients of $U(1) \times SO(\NC)$ and $U(1) \times USp(2\MC)$ for the
end of this section. In order to construct the $SO(\NC),\ USp(2\MC)$ 
hyper-K\"ahler quotient we need to consider ${\cal N}=2$
hypermultiplets. Hence, we consider an ${\cal N}=2$ extension of the
${\cal 
  N}=1$ K\"ahler potential (\ref{eq:so_usp_kahler}), together with
the superpotential 
\beq
\tilde K_{SO,USp} &=& \Tr\left[ QQ^\dagger e^{-V'} + \tilde Q^\dagger \tilde Q e^{V'} + 
\lambda \left(e^{-V'{}^{\rm T}} J e^{-V'} - J \right) \right]\ ,
\label{eq:hyper_kahler_so_usp_1}\\
W &=& \Tr \left[Q\tilde Q \Sigma' + \chi \left(\Sigma'{}^{\rm T}J + J
  \Sigma'\right) \right]\ , 
\label{eq:superpot_so_usp_1}
\eeq
where $(V',\Sigma')$ denote the $SO(\NC)$ or $USp(2\MC)$ vector
multiplets, $(Q,\tilde Q^\dagger)$ are $\NF$ hypermultiplets in the fundamental representation of $SO(\NC)$ or $USp(2\MC)$, and 
$(\lambda,\chi)$ are the Lagrange multipliers which are $\NC$-by-$\NC$ matrix valued superfields. 

We can rewrite the K\"ahler potential (\ref{eq:hyper_kahler_so_usp_1})
as follows 
\beq
\tilde K_{SO,USp} 
= \Tr\left[ QQ^\dagger e^{-V'} + J^{\rm T} e^{-V'} J \tilde Q^{\rm T} \tilde Q^* \right]
= \Tr \left[ {\cal Q} {\cal Q}^\dagger e^{-V'}\right]\ ,\quad
{\cal Q} \equiv \left( Q,\ J\tilde Q^{\rm T}\right)\ ,
\eeq
where we have used $e^{V'{}^{\rm T}} = J^{\rm T} e^{-V'} J$. 
This K\"ahler potential is nothing but the ${\cal N}=1$ K\"ahler
potential of $SO(\NC)$ and $USp(2\MC)$ with ${\cal Q}$, a set of
$2\NF$ chiral superfields. 
We can straightforwardly borrow the result of
Sec.~\ref{sec:kahler_so_usp} and hence the K\"ahler potential reads  
\beq
\tilde K_{SO,USp} = \Tr\left[
\sqrt{{\cal M}{\cal M}^\dagger}
\right]\ ,\qquad
{\cal M} \equiv {\cal Q}^{\rm T} J {\cal Q}\ .
\eeq
The constraint coming from the superpotential
(\ref{eq:superpot_so_usp_1}) is 
\beq
Q \tilde Q J=J\tilde Q^{\rm T}Q^{\rm T}
\quad \Rightarrow \quad
{\cal Q}\tilde J {\cal Q}^{\rm T}=0\ ,
\quad {\rm with~}
\tilde J\equiv 
\left(
\begin{array}{cc}
 {\bf 0}& {\bf 1}_{\NF} \\
-\epsilon {\bf 1}_{\NF} & {\bf 0}
\end{array}
\right)\ .
\eeq
Therefore, we again find the constraints for the meson field ${\cal M}$
\begin{eqnarray}
 {\cal M}^{\rm T}=\epsilon {\cal M}\ ,\quad {\cal M}\tilde J{\cal
   M}=0\ ,\quad
\NC-2< {\rm rank}\,{\cal M}\le \NC\ .
\end{eqnarray} 
As is well-known, the $SO(\NC)$ case has a $USp(2\NF)$ flavor symmetry 
while the $USp(2\MC)$ case has an $O(2\NF)$ flavor symmetry. 
Therefore the $USp(2\NF)$ and $O(2\NF)$ isometries act on 
the $SO(\NC)$ and $USp(2\MC)$ hyper-K\"ahler quotients, respectively. 
The resultant spaces can be written locally in generic points as
\begin{align}
 {\cal M}^{\rm HK}_{SO(\NC)} &\simeq {\mathbb R}_{>0}^{\NC} \times 
 \frac{USp(2\NF)}{USp(2\NF-2\NC)\times (\mathbb{Z}_2)^{\NC-1}}\ 
  \supset {\mathbb R}^{\NC}_{>0} \times 
  \frac{U(\NF)}{U(\NF-\NC)\times (\mathbb{Z}_2)^{\NC-1}}\ ,
  \label{eq:SO-HKQ}\\
 {\cal M}^{\rm HK}_{USp(2\MC)} &\simeq
 {\mathbb R}_{>0}^{\MC} \times 
  \frac{SO(2\NF)}{SO(2\NF-4\MC) \times USp(2)^{\MC} }\ 
 \label{eq:USp-HKQ}
 \supset {\mathbb R}_{>0}^{\MC} \times 
  \frac{U(\NF)}{U(\NF-2\MC) \times USp(2)^{\MC} }\ ,
\end{align}
for the $SO(\NC)$ and $USp(2\MC)$ hyper-K\"ahler quotients, respectively. 
These are hyper-K\"ahler spaces of cohomogeneity $\NC$ and $\MC$, 
respectively.\footnote{
Any smooth hyper-K\"ahler manifold of cohomogeneity one, must be 
the cotangent bundle over the projective space, 
$T^* {\mathbb C}P^{\NF-1}$ or flat space \cite{DS}. 
For the $U(1)$ hyper-K\"ahler quotient with $\NF$ flavors, 
the space is of cohomogeneity one: 
${\mathbb R}_{>0} \times SU(\NF)/SU(\NF-2)$. 
This space is blown up to a smooth manifold 
$T^* {\mathbb C}P^{\NF-1}$ once the FI parameters 
are introduced for the $U(1)$ gauge group. 
The result of Ref.~\cite{DS} implies that 
hyper-K\"ahler spaces of cohomogeneity one in Eqs.~(\ref{eq:SO-HKQ}) and
(\ref{eq:USp-HKQ}) must have a singularity.
}
The right-most ones denote the corresponding $SO(\NC)$ and $USp(2\MC)$ 
K\"ahler quotients given in Eqs.~(\ref{eq:SO-KQ}) and
(\ref{eq:USp-KQ}), respectively.  
These K\"ahler spaces are special Lagrangian subspaces
of the hyper-K\"ahler spaces.
As in the K\"ahler cases (\ref{eq:SO-KQ}) and (\ref{eq:USp-KQ}), 
the isotropy (unbroken flavor symmetry) changes from point to
point. It is enhanced when some eigenvalues coincide.

Let us make a comment on the relation to the instanton moduli space.
In Eq.~(\ref{eq:USp-HKQ})
the simplest case of the $USp(2) \simeq SU(2)$ hyper-K\"ahler quotient 
was previously found in \cite{Antoniadis:1996ra} to be
\begin{eqnarray}
{\cal M}^{\rm HK}_{USp(2)\simeq SU(2)} \simeq 
 {\mathbb R}_{>0} \times 
  \frac{SO(2\NF)}{SO(2\NF-4) \times USp(2) }\  .
\end{eqnarray}
This is a hyper-K\"ahler cone and is particularly important because 
the single instanton moduli space of an $SO(2\NF)$ gauge theory 
is the direct product of this space and ${\mathbb C}^2$ i.e.~the position. 
Here ${\mathbb R}_{>0}$ parametrizes the size while
the coset part parametrizes the orientation of a single BPST instanton 
embedded into the $SO(2\NF)$ gauge group.
The moduli space of $k$ instantons in $SO(\NC)$ and $USp(2\MC)$ gauge theories 
are known to be given by $USp(2k)$ and $O(k)$ hyper-K\"ahler quotients, 
respectively \cite{Christ:1978jy,Dorey:1999pd}. 
Compared with our spaces in Eqs.~(\ref{eq:SO-HKQ}) and (\ref{eq:USp-HKQ}),
the instanton moduli spaces contain adjoint fields of $USp(2k)$ and
$O(k)$ too and thus are larger. 
Inclusion of adjoint fields remains as a difficult but important problem.

\bigskip
Before closing this section 
we make a comment on the hyper-K\"ahler quotient of 
$U(1) \times SO(\NC)$ and $U(1) \times USp(2\MC)$.
We succeeded in constructing the hyper-K\"ahler quotient of $SO(\NC)$ and 
$USp(2\MC)$ thanks to the fact that $J\tilde Q^{\rm T}$ is in the fundamental representation, which is the same representation as
$Q$. Although, we want to make use of the same strategy for 
$U(1) \times SO(\NC)$ and $U(1) \times USp(2\MC)$ as before, $J\tilde
Q^{\rm T}$ still has charge $-1$ with respect to the $U(1)$ gauge
symmetry while 
$Q$ has $U(1)$ charge $+1$. Therefore, it is not easy to construct the 
$U(1) \times SO(\NC)$ and $U(1) \times USp(2\MC)$ and we will not solve
this problem in this article.

\section{1/2 BPS Configurations: NL$\sigma$M Lumps}
\label{sec:nlsmlumps}

In this section we will study NL$\sigma$M lumps which are 1/2
BPS configurations. 
Lumps are stringy topological textures 
extending for instance in the $x^3$ direction 
in $d=1+3$ dimensional spacetime
and are supported by the non-trivial second homotopy group $\pi_2({\cal M})$
associated with a holomorphic map from the 2 dimensional spatial
plane $z=x_1 + i x_2$ to a 2-cycle of the target space of the
NL$\sigma$M. 
We will consider the $\mathbb{C}$-plane together with the point at
infinity, that is $z\in \mathbb{C}\cup \{\infty\}\simeq S^2$, which is
mapped into the target space. 
Lumps in non-supersymmetric $SO(\NC)$ theories were studied
in Ref.~\cite{Benson:1994dp} where the second homotopy group is $\pi_2
     [SU(\NC)/SO(\NC)] \simeq {\mathbb Z}_2$  
and therefore those lumps are non-BPS. 
Here we do not consider this type of lumps. 
We will first study BPS lumps in the NL$\sigma$M of 
$U(1)\times G'$ K\"ahler quotients in general, 
then we investigate lumps in the case of $G'=SO,USp$ which have been
constructed in previous sections.

\subsection{Lumps in $U(1)\times G'$ K\"ahler Quotients}
In the NL$\sigma$M of 
$U(1)\times G'$ K\"ahler quotients,  
(inhomogeneous) complex coordinates $\{\phi^\alpha\}$ 
of the K\"ahler manifold, which are the lowest scalar components of  
the chiral superfields, 
are given by some set of holomorphic $G'$ invariants
$I^i$ modulo $U(1)^{\mathbb C}$, 
namely $\phi^\alpha \in \{I^i\}/\!\!/U(1)^{\mathbb C}$. 
Static lump solutions can be obtained by just imposing $\phi^\alpha$ 
to be a holomorphic function
with respect to $z$ 
\beq
\phi^\alpha(t,z,\bar z,x^3) \to \phi^\alpha (z;\varphi^i)\ ,
\label{eq:lump_sol}
\eeq
where $\varphi^i$ denote complex constants. 
The tension of the lumps can be obtained by
plugging the solution back into the Lagrangian
\beq
T = 2\int d^2x \ K_{\alpha\bar \beta}(\phi,\bar \phi)
~\p \phi^\alpha \bar \p \bar \phi^{\bar \beta}\bigg|_{\phi \to
  \phi(z)} = 2\int d^2 x \ \bar \p \p K(\phi,\bar \phi)\bigg|_{\phi \to
  \phi(z)} \ ,
\eeq
where $K$ is the K\"ahler potential and $K_{\alpha \bar \beta} = \p_\alpha \bar \p_{\bar\beta} K$ is the K\"ahler metric. 
We would like to stress that all the parameters $\varphi^i$ are
nothing but the moduli parameters of the 1/2 BPS lumps. 

We assume that the boundary of $z \rightarrow \infty$ is mapped to 
a single point $\phi^\alpha(z) \rightarrow \phi^\alpha_{\rm vev}$ on
the target space.  
Since the functions $\phi^\alpha(z)$ should be
single valued, $\phi^\alpha(z)$ can be expressed with a finite number
of poles as 
\begin{eqnarray}
 \phi^\alpha(z)=\phi_{\rm
   vev}^\alpha+\sum_{i=1}^k\frac{\phi_i^\alpha}{z-z_i} +
 \mathcal{O}(z^{-2}) \ . 
\end{eqnarray}
Strictly speaking, we have to change patch of the target manifold at 
the poles to describe the solutions correctly. 
To describe the lump solutions, 
it is convenient to use the holomorphic $G'$ invariants 
$I^i$ satisfying the constraints as homogeneous coordinates.  
The holomorphic map is expressed by the homogeneous coordinates
$I^i(z)$ which are holomorphic in $z$  
\begin{eqnarray}
I^i(z)=I_{\rm vev}^i z^{n_i\nu} + {\cal O}(z^{n_i\nu-1})\ , 
\label{eq:lumpsolution}
\end{eqnarray}
where $n_i$ is the $U(1)$ charge of the holomorphic $G'$ invariant $I^i$,
and $\nu$ is some number.
$I_{\rm vev}^i$ denotes the vacuum expectation value of $I^i$ at
spatial infinity.
Since all $n_i\,\nu$ must take value in ${\mathbb Z}_{>0}$, we can
express $\nu = k/n_0$ with the greatest common divisor (GCD) $n_0$ of
$\{n_i\}$ and $k$ a non-negative integer. 
The integer $k$ will be found to be the topological winding number.
These polynomials are basic tools 
to study lump solutions and their moduli, and $\phi^\alpha(z)$ can be
written as ratios of these polynomials, namely $U(1)^{\mathbb C}$ invariants, 
which are known as rational maps in the Abelian case.

There is a remark in store for constructing lump solutions.
If a holomorphic map (\ref{eq:lumpsolution}) touches the unbroken phase
of the original gauge theory at some point, 
the behavior of the lump is ill-defined there in terms of the
NL$\sigma$M.  
Generally speaking, as we will see in examples later, the lump
configuration becomes singular at that point.  
Therefore, we have to exclude such singular configurations and
{\it all points in the base manifold $\mathbb{C}$ must be mapped 
to the full Higgs phase by the holomorphic map (\ref{eq:lumpsolution})}.
We will denote this condition the {\it lump condition}.
In other words, there exist limits where lump configurations become
singular by varying the moduli parameters.
For instance, the invariants $I^i(z)$ are prohibited
from having common zeros by the lump condition.
Since common zeros cannot be detected even in the vicinity of
a corresponding point in the base space,
an emergence of common zeros
indicates a small lump singularity, which is well-known
for lumps in the $\mathbb{C}P^n$ model. The lump condition requires
non-vanishing size moduli there. 
As we will show in examples later,
this situation implies the emergence of a local vortex. The lump
condition is stronger than the condition of no common zeros 
in the invariants, except for the $U(N)$ case \cite{Eto:2007yv},
where in fact both the conditions are equivalent.
The difference between the two conditions above implies the existence
of limits where a lump configuration becomes singular
with a non-vanishing size.
This is a typical property of lumps in a NL$\sigma$M with a singular
submanifold. 
We will see explicit examples of this property later.

\subsection{Lump Moduli Spaces vs.~Vortex Moduli Spaces}
\def\hyphen{\hbox{\scriptsize-}}
As a NL$\sigma$M can be obtained in the strong gauge coupling limit
of the gauge theory, lump solutions in such NL$\sigma$Ms 
can also be given as that limit of semi-local vortex solutions,
whose configuration can smoothly be mapped to the Higgs phase.
Therefore, lump solutions are closely related to semi-local vortices in
the original gauge theory, even with a finite gauge coupling. 
Lumps in the $U(\NC)$ K\"ahler quotient, 
namely in the Grassmann sigma model, have been studied 
previously in Refs.~\cite{Din:1981bx,Shifman:2006kd,Eto:2007yv}.
In fact, the dimensions of both the moduli spaces coincide
${\rm dim}_{\mathbb C} {\cal M}_{U(\NC),\NF}^{k\hyphen{\rm vortex}} 
= {\rm dim}_{\mathbb C} {\cal M}_{U(\NC),\NF}^{k\hyphen{\rm lump}} = k\NF$ \cite{Hanany:2003hp,Eto:2006pg}.
It has been found that the moduli space of $k$ lumps in the Grassmann
sigma model is identical to that of $k$ semi-local vortices with
the lump condition in Ref.~\cite{Eto:2007yv}. 
Hence, the inclusive relation is 
${\cal M}_{U(\NC),\NF}^{k\hyphen{\rm vortex}} 
\supset {\cal M}_{U(\NC),\NF}^{k\hyphen{\rm lump}}$.
The lump condition excludes subspaces of 
${\cal M}_{U(\NC),\NF}^{k\hyphen{\rm vortex}}$
corresponding to the minimal size vortices
whose size is of order of the inverse gauge coupling.

In this section we will discuss the relation between moduli spaces for
lump solutions and vortex solutions in the $U(1)\times SO(\NC)$ and
$U(1)\times USp(2\MC)$ cases.  
Here we take  $\NC=\NF$ and $\det M_{\rm vev}\not=0$ for simplicity.
The dimension of the moduli space of $k$ vortices in a $U(1)\times
G'$ gauge theory ($\NF=\NC$) has been found to be \cite{Eto:2008yi}
\beq
{\rm dim}_{{\mathbb C}} {\cal M}_{U(1)\times G'}^{k\hyphen{\rm vortex}} 
= k \NC^2/n_0,
\label{eq:dim_vor}
\eeq
with $\NC = 2\MC$ for $USp(2\MC)$.
In the following, we will count the dimensions of the lump  moduli
spaces.
(We will use the same characters for lowest scalar 
components of chiral superfields as for the superfields themselves). 

In the $U(1)\times SO(2\MC)$ case $(\NC = 2\MC)$, lump solutions with fixed boundary
 conditions are given by taking the following polynomials as 
the holomorphic invariants $I^i=\{M,B\}$ 
defined in (\ref{eq:meson-baryon}).
Their $U(1)$ charges are $\{2,2\MC\}$, respectively.
Thus, their GCD is $n_0 = 2$ and we find
\begin{eqnarray}
  M(z)=M_{\rm vev} z^k+{\cal O}(z^{k-1})\ ,\quad 
  B(z)=B_{\rm vev} z^{k \MC}+{\cal O}(z^{k\MC-1})\ ,\label{eq:SOevenlumpsol}
\end{eqnarray}
with $k\in {\mathbb Z}_{>0}$. Note that we should not neglect the baryon
field $B$, although the baryon field $B$ is dependent on $M$. 
This is because the baryon field $B$ determined by $M(z)$ is not
necessarily holomorphic everywhere in the complex plane $\mathbb C$: 
\begin{eqnarray}
 \det(J) B(z)^2=\det M(z)\ .\label{eq:lump_constraint}
\end{eqnarray}
Generically, this gives $2k\MC$ constraints for moduli parameters.   
For instance, with a single lump solution in the $U(1)\times SO(2)$ case,
a general form of $M(z)$ is given by setting $M_{\rm vev}=\sigma_1$ and $k=1$ 
\begin{eqnarray}
 M(z)=\left(
\begin{array}{cc}
 b& z-a\\ z-a &c 
\end{array}\right)
\quad \rightarrow\quad  \det M(z)=bc-(z-a)^2 \ .
\end{eqnarray}
The constraint (\ref{eq:lump_constraint})
requires $\det M(z)$ to be exactly a square of a polynomial
and then we find the non-trivial conditions; $b=0$ or $c=0$
where the intersection point $b=c=0$ is excluded by the lump
condition. These two disconnected solutions 
correspond to two different types of lumps wrapping different 
$\mathbb CP^1$'s of $\mathcal M_{\NF=2}^{U(1) \times SO(2)} = \mathbb
CP^1 \times \mathbb CP^1$ in Eq.~(\ref{eq:mod_vac_U(1)xSO(2)}). 
For generic $k$-lump configurations, 
we can count the degrees of freedom of the moduli parameters as 
\begin{eqnarray}
 {\rm dim}_{\mathbb C} \, {\cal M}^{k\hyphen{\rm lump}}_{SO(2\MC)}&=&
\#\hbox{moduli in~}M(z)+\#\hbox{moduli in~}B(z)-\#{\rm constraints}\nn
&=&k \frac{(2\MC)(2\MC+1)}{2}+ k \MC- 2k\MC=2k\MC^2\ . 
\end{eqnarray}
In the $U(1)\times SO(2\MC+1)$ case, the $U(1)$ charges of the
invariants $\{M,B\}$ are $\{2,2\MC +1\}$. Hence their GCD is $n_0 = 1$
and lump solutions are given by the following polynomials
\begin{eqnarray}
  M(z)=M_{\rm vev} z^{2k}+{\cal O}(z^{2k-1})\ ,\quad 
B(z)=B_{\rm vev} z^{(2\MC+1)k}+{\cal O}(z^{(2\MC+1)k-1})\ .
\label{eq:SOoddlumpsol}
\end{eqnarray}
The dimension of the $k$-lump moduli space in this case is generically 
given by
\begin{align}
 {\rm dim}_{\mathbb C} \, {\cal M}^{k\hyphen{\rm lump}}_{SO(2\MC+1)}
=2k \frac{(2\MC+1)(2\MC+2)}{2} + k (2\MC+1)- 2k(2\MC+1)=k(2\MC+1)^2\ . 
\end{align}

These two results are the same as those of the 1/2 BPS vortex moduli
spaces derived from the index theorem \cite{Eto:2008yi}, see
Eq.~(\ref{eq:dim_vor}).  
That is, at least for generic points of the lump moduli space, the
moduli for the lump solutions are sufficient to describe the vortex
moduli space in the original gauge theory, and there are no internal moduli 
unlike the orientational moduli ${\mathbb C}P^{\NC-1}$ of the $U(\NC)$
case with $\NF=\NC$ flavors.
This property is significantly different from the $U(\NC)$ case
with the minimal number of flavors $\NF=\NC$, where only local vortices carrying
the orientational moduli exist and the strong coupling limit of them
are not lumps but singular objects of zero sizes.

In the $U(1)\times USp(2\MC)$ case, the baryon field is completely 
described by the meson fields and there are no constraints
\begin{eqnarray}
 M(z)=M_{\rm vev}z^k+{\cal O}(z^{k-1})\ ,\quad  
 B(z)=({\rm Pf}J)^{-1}{\rm Pf}(M(z))\ .\label{eq:USplumpsol}
\end{eqnarray}
Therefore, the number of complex parameters in $M(z)$
is simply given by
\begin{eqnarray}
\#\hbox{moduli in~}M(z)=k\frac{2\MC(2\MC-1)}2=
{\rm dim}_{\mathbb C}\,{\cal M}_{USp(2\MC)}^{k\hyphen{\rm vortex}}-k\MC\ .
\end{eqnarray}
Note that it is different from the dimensions of
the vortex moduli space. 
This deficit number $\MC$ for each lump 
can be understood as follows. In this case, 
color-flavor symmetries $USp(2)^\MC\simeq SU(2)^\MC$ 
survive even at a generic point in the vacuum as we explained 
below Eq.~(\ref{eq:USpvacuum}). 
These surviving symmetries are broken in a vortex configuration and 
this means that the vortex configuration 
has orientational moduli $({\mathbb C}P^1)^{\MC}$ as NG modes.
These modes are expected to be
localized in the Coulomb phase 
of the original
gauge theory, 
which corresponds to the curvature singularity of the NL$\sigma$M,
 and therefore, cannot be detected as moduli of lump solutions 
in the NL$\sigma$M. 
Therefore, roughly speaking, we guess that
\begin{eqnarray}
{\cal M}_{USp(2\MC)}^{k\hyphen{\rm vortex}} \sim
{\cal M}_{USp(2\MC)}^{k\hyphen{\rm singular\ lump}} 
\times ({\mathbb C}P^1)^{k\MC}\ ,
\end{eqnarray}
where ${\cal M}_{USp(2\MC)}^{k\hyphen{\rm singular\ lump}}$ is the
would-be lump moduli space which is parametrized by the complex
parameters in the meson field $M(z)$. 
Emergence of these internal moduli is strongly related to singular
configurations of lumps.\footnote{
This situation is similar to the case of a $U(\NC)$ gauge theory with
$\NF=\NC$ flavors. The gauge theory has a non-Abelian vortex whose
internal moduli space is ${\mathbb C}P^{\NC-1}$. But the strong gauge
coupling limit yields a NL$\sigma$M of only a point and there are no lump
solutions. 
}
Actually, to get regular solutions from lumps in any NL$\sigma$M, we
have to require the lump condition, which means that the rank of the
meson $M$ should be $2\MC$ everywhere in this $USp(2\MC)$ case.
Therefore, no regular solutions exist in the case of $\NF=2\MC$,
because ${\rm Pf} M$ are polynomials in $z$ with order $\MC k$ and
thus has $k \MC$ zeros.
We will show a concrete example in the next subsection.
We expect that each of the orientational moduli 
${\mathbb C}P^1$ are attached to such zeros and the deficit dimension
of ${\cal M}_{USp(2\MC)}^{k\hyphen{\rm singular\ lump}}$ should be
strongly related to the non-existence of regular solutions. 
Regular lump solutions require the number of flavors to be greater than
$2\MC$.

In both cases of $U(1)\times SO(\NC)$ and $U(1) \times USp(2\MC)$
gauge theories, additional NG zero modes can emerge 
as the moduli of vortex configurations if 
we choose special points as the vacuum, $M_{\rm vev}(B_{\rm vev})$.
Especially, by choosing $M_{\rm vev}=J$ ($\mu_i=1$ for all $i$),
the following moduli spaces for a single local 
vortex were found as 
\cite{Eto:2008yi}
\begin{eqnarray}
 {\cal M}^{\rm vortex}_{G',\,k=1}\supset 
{\cal M}^{\rm local~vortex}_{G',\,k=1}={\mathbb C}\times
\frac{G'}{U(\MC)}\ ,
\quad G'=SO(2\MC),\,USp(2\MC)\ , 
\end{eqnarray}
which cannot be moduli of single lump configurations.

To completely treat the vortex moduli, including internal moduli,
we need to use {\it the moduli matrix formalism} \cite{Eto:2006pg}.
This formalism is obtained by merely rewriting 
the holomorphic gauge invariants $M(z),\,B(z)$ 
in terms of the original chiral field $Q(z)$ 
whose components are also polynomials in the complex coordinate $z$.\footnote{
The way to derive the moduli matrix here is slightly
different from the way used in \cite{Eto:2008yi}.
These two ways can be identified by considering 
BPS vortex solutions in the superfield formulation \cite{Eto:2006uw}.
The key observation is that the gauge symmetry $G$ in the
supersymmetric theory is complexified : $G^{\mathbb C}$. 
Hence, the moduli matrix naturally appears in the superfield
formulation, while if we fix $G^{\mathbb C}$ in the Wess-Zumino gauge,
the scalar field $Q_{\rm wz}$ appears as the usual bosonic component
in the Lagrangian.
The moduli matrix is usually denoted by the symbol $H_0(z)$ in
the literature.
}
The description of the lump solutions with respect to $Q(z)$ is
redundant, since $Q(z)$ and $Q'(z)$ determine the same holomorphic map
$M(z),\,B(z)$, if they are related by a complexified gauge
transformation $Q'(z) = V(z) Q(z)$. 
Therefore we have the following equivalence relation, called the
$V$-equivalence  
\begin{eqnarray}
 Q(z)\sim  V(z)Q(z)\ , \quad 
 V(z) \in U(1)^{\mathbb C} \times SO(\NC)^{\mathbb C},\hs{2} U(1)^{\mathbb C} \times USp(2\MC)^{\mathbb C}\ .
\end{eqnarray}
The parameters contained in $Q(z)$ after gauge fixing,
parametrize the moduli space of vortices.
Conversely, all moduli of vortices including internal moduli 
are contained in $Q(z)$, and thus $Q(z)$ is denoted
{\it the moduli matrix}. In this formalism 
the boundary conditions 
(\ref{eq:SOevenlumpsol}), (\ref{eq:SOoddlumpsol}) and 
(\ref{eq:USplumpsol}) are interpreted as
 constraints for the moduli matrix $Q(z)$ \cite{Eto:2008yi}
\begin{eqnarray}
&SO(2\MC), \,USp(2\MC):&
Q^{\rm T}(z)JQ(z)=M_{\rm vev}z^{k}+{\cal O}(z^{k-1})\ ,\nn
&SO(2\MC+1):&
Q^{\rm T}(z)JQ(z)=M_{\rm vev}z^{2k}+{\cal O}(z^{2k-1})\ .
\label{eq:Qzconstraint}
\end{eqnarray}
The constraint (\ref{eq:lump_constraint}) is of course automatically
solved in this formalism. 
This formalism is apparently independent of the gauge coupling and it
is well-defined to require the lump conditions to hold on the vortex
moduli space.
We expect that a submanifold of the $k$-vortex moduli space satisfying
the lump condition is equivalent to the $k$-lump moduli space,
\begin{eqnarray}
 {\cal M}^{k\hyphen{\rm lump}}\simeq 
\left\{a | a\in {\cal M}^{k\hyphen{\rm vortex}}, 
~\hbox{the lump condition}\right\}.
\end{eqnarray} 
This expectation is quite natural and is enforced by the above
observations by counting the dimensions. 
Because, if we can consider a NL$\sigma$M as an approximation to the
gauge theory with a strong but finite gauge coupling $g$, a lump
solution should describe an approximate configuration of a vortex,
whereas a steep configuration with a width of order $1/g\sqrt{\xi}$ is
excluded by some UV cutoff $\Lambda< g\sqrt{\xi}$.  
Of course, to justify this expectation, we need to verify an 
equivalence\footnote{
In the $U(1)\times USp$ and $U(1)\times SO$ cases, we have to verify
that the meson field $M(z)$ whose elements are polynomials can be
always decomposed in $Q(z)$ whose elements are also polynomials and
furthermore that there is no degeneracy of moduli in the construction
of $M(z)$ from $Q(z)$ under the lump condition.
There is no known proof and it is expected to be technically
complicated. 
} 
between the two formalisms, the moduli matrix formalism and the
holomorphic map (\ref{eq:lumpsolution}) with the constraint on the
invariants, under the lump condition.  
In examples of the next subsection, we just assume that this
expectation is true.
To construct lump solutions for large $\NF(\NC)$,  the moduli matrix
formalism is somewhat easier than treating $M(z),B(z)$ as they are.

\subsection{Lumps in $U(1)\times SO(2\MC)$ and $U(1)\times USp(2\MC)$ 
K\"ahler Quotients}

\subsubsection{BPS Lumps in the $U(1)\times SO(2\MC)$ K\"ahler Quotient}

Let us start with the simplest example in which the gauge group is
$U(1) \times SO(2)$ with two flavors $\NF=2$. As we have
studied in Sec.~\ref{sec:example}, the target space is ${\mathbb
  C}P^1 \times {\mathbb C}P^1$. 
Lump solutions are classified by a pair of integers $(k_+,k_-)$ 
given as 
\beq
\pi_2\left({\cal M}_{\NF=2}^{U(1)\times SO(2)}\right) = {\mathbb Z}
\times {\mathbb Z} \ni (k_+,k_-)\ .
\eeq
A solution with $(k_+,k_-)$ lumps is given by
\beq
Q(z) = 
\left(
\begin{array}{cc}
Q_{1}^+(z) & Q_{2}^+(z)\\
Q_{1}^-(z) & Q_{2}^-(z)
\end{array}
\right)\ ,
\eeq
where $Q_{+i}(z), Q_{-i}(z)$ are holomorphic functions of $z$ of
degree $k_{\pm}$, respectively. 
One can verify that the tension is given by
\beq
T =  \int d^2x \ 2 \p \bar\p K_{U(1)\times SO(2)} = \pi \xi (k_+ +
k_-) \equiv \pi \xi k\ , 
\eeq
where $K_{U(1)\times SO(2)}$ is the K\"ahler potential 
given in Eq.~(\ref{eq:kahler_SO(2)}). 
Interestingly, the tension of the minimal lump $(k_+,k_-)=(1,0),(0,1)$ 
is half of $2\pi\xi$ which is that of the minimal lump in the 
usual ${\mathbb C}P^1$ model. A similar observation has been obtained
recently in Ref.~\cite{Eto:2008yi}.

Next, we would like to consider lump configurations in slightly
more complicated models by considering general $U(1)\times SO(2\MC)$ K\"ahler
quotients, where we set $\MC \ge 2$, $\NF = 2\MC$ and $M_{\rm vev}=J$. 
As an example for $k=1$, we take
\beq
Q_{k=1} = 
\left(
\begin{array}{cc}
z {\bf 1}_\MC - A & C\\
0 & {\bf 1}_\MC
\end{array}
\right)\ ,
\qquad
\left\{
\begin{array}{c}
A = {\rm diag}(z_1,z_2,\cdots,z_\MC)\ ,\\
C = {\rm diag}(c_1,c_2,\cdots,c_\MC)\ .
\end{array}
\right.
\eeq
These diagonal choices allow us to treat the invariants as if they
were independent invariants of $\MC$ different $SO(2)$'s. 
Hence, one can easily find an $SO(2)$ part inside $M$ as
\begin{eqnarray}
 \left(
\begin{array}{cc}
(M)_{i,i} & ( M)_{i,i+\MC}\\
( M)_{i+\MC,i} & ( M)_{i+\MC,i+\MC}
\end{array}
\right)
=
\left(
\begin{array}{cc}
0 & z-z_i\\
 z - z_i & 2c_i
\end{array}
\right)\ ,\quad i = 1,2,\cdots,\MC\ ,\label{eq:SOmesonsol}
\end{eqnarray} 
which satisfies the constraint (\ref{eq:Qzconstraint}).
Note that 
non-zero parameters $c_i$ keep the ${\rm rank} \, M \ge 2\MC-1$, even at
$z=z_i$.  
All their eigenvalues are also eigenvalues of $ M M^\dagger$
\beq
\lambda_{i\pm} = |z-z_i|^2 + 2|c_i|^2 \pm 2|c_i|\sqrt{|z-z_i|^2 +
  |c_i|^2}\ .
\eeq
Thus, the K\"ahler potential in Eq.~(\ref{eq:Kahler_pot_UxSO}) becomes
\beq
K = \xi \log \left[ \sum_{i=1}^M \left( \sqrt{\lambda_{i+}} +
  \sqrt{\lambda_{i-}} \right) \right] 
= \xi \log \left( 2 \sum_{i=1}^M \sqrt{ |z-z_i|^2 + |c_i|^2 } \right)\
. \eeq
The energy density is obtained by 
$\mathcal{E}=2\partial\bar \partial K$ with this K\"ahler potential 
and exhibits an interesting structure. It is
proportional to the logarithm of the sum of the square root of $|P_i(z)|^2$,
while the known K\"ahler potential of a ${\mathbb C}P^M$ lump is just
the logarithm of the sum of $|P_i(z)|^2$. 
This difference gives us quite distinct configurations.
If we take some $c_i$ to vanish, then we find that the energy density
of the configuration becomes singular at $z=z_i$
\begin{eqnarray}
\mathcal{E} = 2\xi\partial\bar \partial\log\left(\sqrt{|z-z_i|^2}
+\cdots\right)\sim 
{\rm const.}\times \frac{1}{|z-z_i|}+{\cal O}(z^0)\ .
\end{eqnarray} 
This is due to the curvature singularity which appears when the
manifold becomes of ${\rm rank}\,M=2\MC-2$, and in other words, 
violate the lump condition.
Note that this singular configuration has a non-vanishing size, as we
mentioned above.
If we take all $z_i$'s and all $c_i$'s to be coincident, respectively,
we find that the K\"ahler potential reduces to that of the minimal
winding one in the $U(1)\times SO(2)$ model. 
This suggests that the trace part of $C$ determines the overall size 
of the configuration and 
the trace part of $A$ corresponds to the center of mass.
As we will explain later, 
only this trace part of $A$ among the parameters 
is a normalizable mode in the effective action of the lump.  

A single lump in $U(1) \times SO(2\MC+1)$ might be almost the same as the
coincident $k=2$ lumps in $SO(2\MC)$. 
However we will not discuss this case in detail.

\subsubsection{BPS Lumps in the $U(1) \times USp(2\MC)$ K\"ahler Quotient}

Let us first examine a lump solution in the $U(1)\times
USp(2)$ theory with $\NF=2$. 
In this case, however, we 
obtain only local vortices and cannot 
observe regular lumps in the NL$\sigma$M
since the vacuum is just a point.
 After fixing the gauge, the chiral field can be expressed as 
\beq
Q(z) = 
\left(
\begin{array}{cc}
z-a & 0\\
b & 1
\end{array}
\right)\ .
\eeq
This matrix yields 
\beq
M= (z-a)\,J\ ,\qquad
K = \frac{\xi}{2}\log|z-a|^2\ .
\eeq
At the center of the vortex, the rank of $M$ always 
reduces to zero, where the $U(1)$ gauge symmetry is restored.
Therefore, solutions are always singular 
at that point, because we know that $USp(2) \simeq SU(2)$ and the $U(2)$
model with 2 flavors admits only local vortices rather than semi-local
vortices which reduce to lumps in the NL$\sigma$M limit. 
Indeed, the parameter $b$ which does not appear in $M$ is the
orientational modulus of local vortex in the original $U(1)\times
USp(2)$ gauge theory and describes ${\mathbb C}P^1$. 

As we have mentioned, 
lump solutions in the case of $\MC=\MF$ always have singular points 
in the configurations.
The simplest non-trivial example for a regular lump
is obtained in the case of $U(1) \times USp(4)$ with $6$ flavors. 
A lump (vortex) solution in this case, with the minimal winding ($k=1$) has 
$\MC\NF=12$ complex parameters.
Let us consider the following field configuration as a typical minimal
example of $k=1$;  
\beq
Q(z) = 
\left(
\begin{array}{cccccc}
z-z_+ & 0 & 0 & c&a_+&0\\
0 & z-z_- & -c & 0&0&a_-\\
0 & 0 & 1 & 0&0&0\\
0 & 0 & 0 & 1&0&0
\end{array}
\right)\ , 
\eeq
which gives the following characteristic polynomial
\begin{eqnarray}
 \det(\lambda-MM^\dagger)=
\lambda^2
\left(\lambda^2-(R_+^2+R_-^2+4|c|^2)\lambda+R_+^2R_-^2\right)^2 \ ,
\end{eqnarray}
with $R_\pm=\sqrt{|z-z_\pm|^2+|a_\pm|^2}$.
Then the energy density of the configuration ${\cal E}$ is given by
\beq
{\cal E}=
2\partial \bar \partial K_{U(1)\times USp(4)}|_{\rm sol}
=\xi \partial \bar \partial
\log\left((R_++R_-)^2+4|c|^2\right) \ . 
\label{eq:lump_USp(4)}
\eeq
This configuration is regular everywhere as long as $a_\pm\not=0$,
that is, it satisfies the lump condition.
If we choose $a_+ = a_-$ and $z_+=z_-$,  it corresponds to a 
${\mathbb C}P^2$ single lump solution.

\subsection{Effective Action of Lumps}
Now we have a great advantage thanks to the above superfield
formulation of the NL$\sigma$M. A supersymmetric low energy
effective theory on the 1/2 BPS lumps is immediately obtained merely
by plugging the 1/2 BPS solution (\ref{eq:lump_sol}) into the K\"ahler
potential which we have obtained in the previous section after
promoting the moduli parameters $\varphi$ to fields on the lump
world-volume 
\beq
\phi^\alpha(t,z,\bar z,x^3) \to \phi^\alpha(z;\varphi^i(t,x^3))\ .
\eeq
The resulting (effective) expression for the K\"ahler potential is
\beq
{\cal K}_{\rm lump} = \int dzd\bar z\  
K\left(\phi(z,\varphi^i(t,x^3),\ \phi^\dagger(\bar
z,\bar\varphi^i(t,x^3)\right)\ .\label{eq:Kahletpot_lump}
\eeq

Let us make a simple example of the ${\mathbb C}P^1$ sigma model which
is the strong coupling limit of a $U(1)$ gauge theory with $\NF=2$
flavors $Q=(Q_1,\ Q_2)$. In this case, $Q_1$ and $Q_2$ themselves 
play the role of the holomorphic invariants $I^i$ 
and the inhomogeneous coordinate is given by $\phi=Q_2/Q_1$. 
We fix the $U(1)^{\mathbb C}$ symmetry in such a way that $Q$ is expressed by 
\beq
Q = (1,\ b)\ .
\eeq
From Eq.~(\ref{eq:Kahler_U(N)_N=1}), the K\"ahler potential and the
corresponding Lagrangian are of the form 
\beq
K = \xi \log (1 + |b|^2)\ ,\qquad
{\cal L} = \xi \frac{|\p_\mu b|^2}{(1+|b|^2)^2} \ .
\eeq
A single 1/2 BPS lump solution in this model is given by
\beq
Q(z) = ( z - z_0, \ a) \hs{3} \leftrightarrow \hs{3} \phi = \frac{a}{z-z_0},
\eeq
where $z_0$ corresponds to the position of the lump and 
$a$ is its transverse size and phase moduli.
To obtain the effective theory of the lump, one needs to promote the
moduli matrix as follows 
\beq
Q(z) = (z - z_0,\ a)
\quad \to \quad
Q(t,z) = (z - z_0(t),\ a(t))\ .
\eeq
Plugging this into the formal expression (\ref{eq:Kahletpot_lump}), 
we get the effective theory
\begin{align}
{\cal L}^{\rm eff} &= \xi \int dz d\bar z\ \delta^t\delta_t^\dagger \log\left(|z-z_0(t)|^2 + |a(t)|^2\right)
\nonumber\\
&= \xi \int dzd\bar z\ 
\left[
\frac{|a(t)|^2}{\left(|z - z_0(t)|^2 + |a(t)|^2\right)^2} \left|\dot z_0 (t) \right|^2 + 
\frac{|z - z_0(t)|^2}{\left(|z - z_0(t)|^2 + |a(t)|^2\right)^2} \left|\dot a (t)\right|^2
\right]\ .
\end{align}
The second term in the second line does not converge, thus the size
modulus $a(t)$ is not dynamical. Hence, we should fix it by hand as
$a(t) = {\rm const} \neq 0$.  
Then the only dynamical field is the translation $z_0(t)$ and the
effective action is 
\beq
{\cal L}_\infty^{\rm eff} = \pi \xi |\dot z_0 (t)|^2\ ,
\eeq
where $2\pi\xi$ is the tension of the minimal winding solution.

\subsection{Identifying Non-normalizable Modes}

We can determine
which parameters in $Q(z)$ are localized on lumps and normalizable, 
and which parameters are non-normalizable.
If there exists a divergence in the K\"ahler potential 
which cannot be removed by the K\"ahler transformations,
it indicates that the 
moduli parameters included in the divergent terms are non-normalizable.
Let us substitute an expansion of the lump solution with respect to $z^{-1}$
\begin{eqnarray}
\phi^\alpha(z) = \phi^\alpha_{\rm vev} + \frac{\chi^\alpha}{z} 
+ {\cal O}(z^{-2}) \ , \quad \chi^\alpha = \sum_{i=1}^k \phi_i^\alpha \ ,
\end{eqnarray}
into the K\"ahler potential (\ref{eq:Kahletpot_lump})
and expand it as well
\begin{eqnarray}
{\cal K}_{\rm lump} &=& \lim_{L \to \infty} \int_{|z| \le L} d^2x \ 
\left[
K(\phi^\alpha_{\rm vev}, \bar \phi^{\bar \beta}_{\rm vev}) 
+ \frac{1}{z} \p_\alpha K \chi^\alpha 
+ \frac{1}{\bar z} \bar \p_{\bar \alpha} K \bar \chi^{\bar \alpha}
+ \frac{1}{|z|^2} \p_\alpha \bar \p_{\bar \beta} K \chi^\alpha \bar \chi^{\bar \beta} + {\cal O}(|z|^{-3})
\right] \nonumber \\
&=& \lim_{L \to \infty} \left[
2 \pi L^2 \ K(\phi_{\rm vev},\bar \phi_{\rm vev})
 + 2 \pi \log L \ \p_\alpha \bar \p_{\bar \beta} K(\phi_{\rm vev},\bar \phi_{\rm vev}) \chi^\alpha \bar \chi^{\bar \beta} + {\cal O}(1)
\right]\ , 
\end{eqnarray}
where $L$ is an infrared cutoff.
Thus we can conclude that the moduli parameters included in
$\{\phi^\alpha_{\rm vev},\chi^\alpha\}$ are all non-normalizable
and the others are normalizable. The modulus $a$ in the last 
subsection is a typical example of $\chi^\alpha$.

For instance, let us take a look at the example (\ref{eq:SOmesonsol})
of the solution for single lumps in the $U(1)\times SO(2\MC)$ case. 
The meson field $M(z)$ 
has the following elements : $(z-z_i)$ and $2 c_i$.
One can partly 
construct inhomogeneous coordinates of the manifold in this case 
by taking ratios from pairs of the elements, 
\beq
\phi^i &=& \frac{2c_i}{z-z_{\MC}} = \frac{2c_i}{z} + {\cal O}(z^{-2}),
\quad {\rm for~} 1\le i\le \MC\ ,\nn
\phi^{i+\MC} &=& 
\frac{z-z_i}{z-z_{\MC}} = 
1 - \frac{z_i - z_{\MC}}{z} + {\cal O}(z^{-2}),
\quad {\rm for~} 1\le i\le \MC-1\ .
\eeq
Thus the moduli $c_i$ and $z_i - z_{\MC}$ are non-normalizable. 
The only normalizable modulus is $\sum_{i=1}^{\MC}z_i/\MC$ 
which is the center of mass. 
This fact is a result of the K\"ahler metric
(\ref{eq:Kahlermet_U1SOUSp}) where 
the trace part of the meson field $M$ does not contribute to the metric. 
Generally speaking, all moduli of a single lump 
in the $U(1)\times SO(2\MC)$ and $U(1)\times USp(2\MC)$ theories are
non-normalizable except 
for the center of mass and 
the orientational moduli of local vortex.

\section{Conclusion and Discussion}
\label{sec:concdisc}

We have explicitly constructed the K\"ahler potentials for
NL$\sigma$Ms describing the Higgs phase of ${\cal N} =1$
supersymmetric $SO(\NC)$ and $USp(2\MC)$ gauge theories.
The key point in the construction lies in the
use of taking the gauge symmetry to be $U(\NC)$ and restricting the
algebra down to either $\mathfrak{so}(\NC)$ or $\mathfrak{usp}(2\MC)$
with Lagrange multipliers.
The result is written both in terms of the component fields and the
holomorphic invariants, i.e.~the mesons and the baryons of the
theories. Because the obtained result is difficult to manage in
practice in the large $\NC$ $(\NF)$ limit, we have developed an
expansion around the vacuum expectation values of the meson field, and
obtained the scalar curvature of both theories, i.e.~$SO(\NC)$ and
$USp(2\MC)$.  
Furthermore, have made the same considerations for the case of
$U(1)\times SO(\NC)$ and $U(1)\times USp(2\MC)$, and obtained the
K\"ahler potential, metric, expansion and curvature also in these
cases. 

Following the same strategy as in the K\"ahler quotient case, we have
been able to obtain the hyper-K\"ahler quotient in the case of
$SO(\NC)$ and $USp(2\MC)$ gauge theories, simply by rewriting the
fields by means of the algebra to fields with $2\NF$ flavors all in
the fundamental representation and we confirm the flavor symmetry of
the $SO(\NC)$ hyper-K\"ahler quotient to be $USp(2\NF)$ and for
$USp(2\MC)$ it is $O(2\NF)$.

A significant feature of those NL$\sigma$Ms, 
is that a point in the target space can reach within a finite distance 
submanifolds corresponding to unbroken phases of the gauge theories. 
We have observed that a curvature singularity emerges there.
If we consider a generic gauge group with a generic representation as
the original gauge theory, we can observe such singularities in many
NL$\sigma$Ms unlike the well-known $U(N)$ (Grassmannian) case.
The NL$\sigma$Ms we have considered here can be regarded as test cases
for those theories.

In the second part of the paper we have studied the $1/2$ BPS,
NL$\sigma$M lumps in $U(1)\times G'$ gauge theories
and observed that we can construct lump solutions straightforwardly if
the K\"ahler potential for the NL$\sigma$M is given in terms of
holomorphic invariants of $G'$. 
We found that counting the dimension of these (regular) lump moduli
spaces gives the same result for the semi-local vortex moduli space
in the case of $SO(\NC)$ and $USp(2\MC)$ theories. 
This fact enforces our natural expectation that
those moduli spaces are homeomorphic to each other 
except in the subspaces where the lump condition is violated.
Furthermore, by considering effective actions within our formalism for
the NL$\sigma$M lumps, we have obtained a conventional method to
clarify the non-normalizability of the moduli parameters in general
cases. 
By using this, we can conclude that in both the cases of
 $U(1)\times SO(2\MC)$ and $U(1)\times USp(2\MC)$ K\"ahler
quotients, all moduli parameters of a single regular lump are
non-normalizable except for the center of mass.

An important observation of lump configurations in $U(1)\times
SO(\NC)$ and $U(1)\times USp(2\MC)$ theories is the existence of
a singularity in the target manifold.
In those theories, a lump configuration becomes singular without taking
the zero size limit, simply if the configuration touches the
singularity of the manifold, whereas a lump in the $U(N)$ case is 
always regular with a finite size and becomes singular only in the zero
size limit.  
Especially, in the case of $U(1)\times USp(2\MC)$ with $\NF=2\MC$,
only singular solutions (with a finite or zero size) exist.

It is an important problem to determine the second homotopy group 
$\pi_2({\cal M}_{U(1)\times (SO,USp)})$ in the case of $U(1)\times
SO(\NC)$ and $U(1)\times USp(2\MC)$ theories.
To support stability of lumps in those models, we expect that 
\begin{eqnarray}
 \pi_2({\cal M}_{U(1)\times SO(\NC)})\simeq \mathbb Z\times \mathbb Z_2\ ,
\qquad \pi_2({\cal M}_{U(1)\times USp(2\MC)})\simeq \mathbb Z\ ,
\end{eqnarray}
where the $\mathbb{Z}_2$ charge for the $U(1)\times SO(\NC)$ case is 
naturally expected, since the corresponding local vortices have their
charges due to  
$\pi_1(U(1)\times SO(\NC)/\mathbb Z_2)=\mathbb Z\times \mathbb Z_2$
\cite{Ferretti:2007rp}. To determine the homotopy group in these cases
is a complicated task since we have to take non-trivial directions of
cohomogeneity into account, and a further study of the moduli space of
lumps beyond counting dimensions also is needed. 
This problem still remains as a future problem.   
The relation between our solutions and the lumps 
in non-supersymmetric $SO(N)$ QCD \cite{Benson:1994dp} 
is, therefore, unclear so far. 
In their case, the lumps are supported by the homotopy 
group $\pi_2 [SU(N_{\rm F})/SO(N_{\rm F})] \simeq {\mathbb Z}_2$. 
Therefore, these lumps are non-BPS. 
In our case, the gauge coupling constants for 
$SO(N_{\rm C})$ and $U(1)$ could be different 
although we did not consider it. 
Let $g$ and $e$ be the gauge couplings of 
the $SO(N_{\rm C})$ and $U(1)$ gauge groups, respectively. 
We have taken the strong gauge coupling limit for 
both the couplings, $g,e \to \infty$, in which case 
the gauge theory reduces to the NL$\sigma$M of 
the $U(1)\times SO(N_{\rm C})$ K\"ahler quotient. 
Without taking the strong coupling limit for $e$, 
the size (width) $1/e \sqrt \xi$ for the ``Abelian''
vortices becomes larger as the $U(1)$ gauge coupling 
$e$ becomes smaller. 
In the limit of vanishing $e$, 
we expect that they disappear and 
only non-BPS ${\mathbb Z}_2$ lumps remain. 
It is important to clarify this point
which also remains as a future problem.

Besides these problems,  
there are many interesting future problems in the following.

In certain models it has been proposed that 
the moduli space of vacua admits a Ricci-flat 
(non-compact Calabi-Yau) metric \cite{Hanany:2008kn}. 
In the case of the $SU(N_{\rm C})$ K\"ahler quotient,
a Ricci-flat metric was obtained by 
deforming the K\"ahler potential (\ref{eq:Kahler_SU}) of 
the original $SU(N_{\rm C})$ gauge theory to 
$K = f\left(\Tr [QQ^\dagger e^{-V'}]\right)$ 
with an unknown function $f$,  
and solving the Ricci-flat condition (the Monge-Ampere equation) 
for $f$ \cite{Higashijima:2001vk}. 
The metric turns out to be the canonical line bundle over the
Grassmann manifold $Gr_{N_{\rm F},N_{\rm C}}$
\cite{Higashijima:2002px}.
It is certainly worthwhile to construct a Ricci-flat metric also on
the $SO$ and $USp$ K\"ahler quotients. 
The expansion (\ref{ExpKahlerSO}) should be enough to determine
the unknown function $f$ with a K\"ahler potential
$K=f\left(\Tr[\sqrt{MM^\dagger}]\right)$.

An extension to hyper-K\"ahler quotients with other gauge groups, 
namely exceptional groups is also an interesting future problem. 
As in Eq.~(\ref{eq:so_usp_kahler}) for $SO(\NC)$ and $USp(2\MC)$
K\"ahler quotients,  K\"ahler quotients may be achieved by introducing
a proper constraint. For instance for a $E_6$ quotient,  
$\Gamma_{ijk} {{\left(e^{V'}\right)}^i}_l {{\left(e^{V'}\right)}^j}_m
{{\left(e^{V'}\right)}^k}_n - \Gamma_{lmn}=0$  
is a candidate constraint to embed $E_6$ into $U(27)$, 
where $\Gamma_{ijk}$ is the third-rank invariant symmetric tensor of $E_6$.
This will be achieved by introducing a Lagrange multiplier 
$\lambda^{lmn}$ belonging to the rank-3 anti-symmetric
representation.
Since the study of vortices in $U(1)\times G'$ with $G'$ being
exceptional groups has been raised in \cite{Eto:2008yi}, lumps in 
these K\"ahler quotients are also interesting subjects to be studied.

We should also consider hyper-K\"ahler quotients for
other representations. 
In particular, including adjoint fields into our work 
is important because the resultant spaces appear as 
multi-instanton moduli spaces of $SO(\NC)$ and $USp(2\MC)$ gauge theories.

In the case of the ${\cal N}=2$ hyper-K\"ahler NL$\sigma$M, 
the only possible potential consistent with eight supercharges is 
written as the square of a tri-holomorphic Killing vector 
\cite{AlvarezGaume:1983ab}.
The explicit potentials can be found for instance for 
$T^* {\mathbb C}P^{N-1}$ \cite{Gauntlett:2000ib,Arai:2002xa}, 
toric hyper-K\"ahler manifolds \cite{Gauntlett:2000bd}, 
$T^* Gr_{N,M}$ \cite{Arai:2003tc} and $T^* F_n$ \cite{Eto:2005wf}.
In terms of the hyper-K\"ahler quotients these potentials are obtained
as usual masses of hypermultiplets in the corresponding ${\cal N}=2$
supersymmetric gauge theories \cite{Arai:2003tc}.  
For this massive deformed hyper-K\"ahler NL$\sigma$M
one can construct domain walls which are the other fundamental 
1/2 BPS objects; 
1/2 BPS domain wall solutions in the
$U(\NC)$ hyper-K\"ahler quotient, 
namely $T^* Gr_{N,\NC}$, see \cite{Isozumi:2004vg}.
Constructing a massive deformation and domain wall solutions 
in $U(1)\times SO(\NC)$ and $U(1)\times USp(2\MC)$ 
hyper-K\"ahler quotients remains as future problems.

Time-dependent stationary solutions, 
called Q-lumps \cite{Leese:1991hr}, 
are also BPS states in a NL$\sigma$M with a potential. 
Q-lumps were constructed in 
the ${\mathbb C}P^1$ model \cite{Leese:1991hr}, 
the Grassmann sigma model ($U(N_{\rm C})$ K\"ahler quotient)
\cite{Bak:2006qk}, 
and the asymptotically Euclidean spaces \cite{Naganuma:2001pu}. 
It is one of the possible extensions to construct Q-lumps in
$U(1)\times SO(N_{\rm C})$ and 
$U(1)\times USp(2M_{\rm C})$ K\"ahler quotients.

As mentioned in the introduction, quantum corrections to the ${\cal
  N}=1$ K\"ahler potentials are also an important and interesting
future direction to follow up on. 

Finally, many extensions and applications of the present works 
include: dynamics of lumps \cite{Ward:1985ij}, cosmic lump strings
\cite{Benson:1993at,Dasgupta:2007ds,Vachaspati:1991dz}
and especially their reconnection \cite{Eto:2006db}, 
composite states like 
triple lump-string intersections \cite{Naganuma:2001pu} 
and lump-strings stretched between domain walls \cite{Isozumi:2004vg}, 
and the Seiberg-like duality \cite{Eto:2007yv}.

\section*{Acknowledgments}
We are grateful to Kenichi Konishi, Takayuki Nagashima and especially to Walter Vinci for fruitful
discussions.  
T.F, M.N and K.O would like to thank the theoretical HEP group at
University of Pisa for their hospitality.  
M.E and S.B.G thank the organizers of the conference ``Continuous Advances in QCD 2008'' for
warm hospitality.
The work of M.E.~and K.O.~(T.F.) is supported by the Research 
Fellowships of the Japan Society for 
the Promotion of Science for Research Abroad (for Young
Scientists).
The work of M.N. is supported in part by Grant-in-Aid for Scientific
Research (No.~20740141) from the Ministry
of Education, Culture, Sports, Science and Technology-Japan.

\appendix
\section{Various Theorems and Their Proofs}
\label{appvartheorem}
\subsection{$SO(2\MC), USp(2\MC)$ Groups and Their Invariant Tensors}
\label{sec:InvariantTensor}
Let us define the following sets of $n$-by-$n$ matrices for $\epsilon=\pm 1$
\begin{eqnarray}
{\rm Inv}_\epsilon(n) \equiv
\{J \, | \, J^{\rm T} = \epsilon J,\ J^\dagger J={\bf 1}_n\}\ .   
\end{eqnarray}
That is, elements of ${\rm Inv}_\epsilon(n)$ 
are (anti)symmetric and unitary.\\ 
{\bf Proposition:}
For arbitrary $A\in {\rm Inv_+(2)}$, there exists a $2$-by-$2$ unitary matrix $u$ such that
\begin{eqnarray}
 A=u^{\rm T} u \ .
\label{eq:proposition}
\end{eqnarray}
{\bf Proof:}
A general solution of $A$ is given by 
\begin{eqnarray}
 A&=&e^{i\lambda}\left(
\begin{array}{cc}
e^{i\rho} \cos \theta& i\sin\theta \\ 
i\sin\theta &e^{-i\rho}\cos\theta
\end{array}\right)\nonumber \\
&=& e^{\frac{i}2(\lambda {\bf 1}_2+\rho\, \sigma_3)}
\left(\cos\theta{\bf 1}_2+i\sigma_1 \sin\theta\right)
e^{\frac{i}2(\lambda {\bf 1}_2+\rho\, \sigma_3)}
= u^{\rm T}u\ , 
\end{eqnarray}
with $u=e^{\frac{i}2 \theta \sigma_1}
e^{\frac{i}2(\lambda {\bf 1}_2+\rho\, \sigma_3)}$
$\in U(2)$. {\tiny $\blacksquare$}

\noindent{\bf Theorem 1-s:}
An arbitrary $A\in {\rm Inv}_+(n)$ can be written as
\begin{eqnarray}
 A=u^{\rm T} u \ , 
\label{eq:O}
\end{eqnarray}
with an $n$-by-$n$ unitary matrix $u$.
{\tiny $\blacksquare$}\\
Therefore we find,
\begin{eqnarray} 
{\rm Inv}_+(n) \simeq U(n)/O(n)\ .
\end{eqnarray}\\
{\bf Proof 1-s:}
It is easy to show that an arbitrary symmetric matrix can be rewritten as 
\begin{eqnarray}
A ~\rightarrow~ A' ~=~ u'A {u'}^{\rm T} ~=~ \left(
\begin{array}{ccccc}
|a_1| & b_1 & 0 & 0 & \cdots \\ 
b_1 & |a_2| & b_2 & 0 & \cdots \\
0 & b_2 & \ddots & \ddots & \\
0 & 0 & \ddots \\
\vdots&\vdots
\end{array}\right) ~\in~ {\rm Inv}_+(n) \ , 
\end{eqnarray}
with an unitary matrix $u'$.
The matrix $A'$ is also a unitary matrix and 
this fact leads to  $b_1=0$ or $b_2=0$. Therefore
\begin{eqnarray}
 A'=\left(
\begin{array}{cc}
 1&{\bf 0} \\ {\bf 0} & A_{(n-1)}
\end{array}\right)\ ,\quad
{\rm or~}\quad
\left(
\begin{array}{cc}
 A_{(2)}&{\bf 0} \\ {\bf 0} & A_{(n-2)}
\end{array}
\right) \ ,
\end{eqnarray}
where $A_{(m)}\in {\rm Inv}_+(m)$.
Recursively, we find $A'$ takes a block-diagonal form which diagonal
elements are $1$ or $2$-by-$2$ symmetric unitary matrices.
By using Proposition\,(\ref{eq:proposition}), we can show that there exists a unitary matrix $\tilde u$ such that $\tilde u A \tilde u^{\rm T} = \mathbf 1_n$, that is, there exists a unitary matrix $u$ such that $A = u^{\rm T} u$. {\tiny $\blacksquare$}
 
By using a similar algorithm, we can show that\\
{\bf Theorem 1-a:}
An arbitrary $A\in {\rm Inv}_-(2m)$ can be rewritten as
\begin{eqnarray}
 A=u^{\rm T}J_{m}^- u, \quad J_m^-=
\left(\begin{array}{cc}
0 & 1 \\ -1 & 0
\end{array}\right)
\otimes {\bf 1}_m \ , 
\label{eq:USp}
\end{eqnarray}
with an appropriate unitary matrix, $u$, ($uu^\dagger={\bf 1}_{2m}$).
{\tiny $\blacksquare$}\\
Therefore we find 
\begin{eqnarray}
{\rm Inv}_-(2m)\simeq U(2m)/USp(2m)\ .
\end{eqnarray} 
A choice of $J_\epsilon \in {\rm Inv}_\epsilon(n)$
defines a subgroup $G_\epsilon(J_\epsilon)$ of $U(n)$ as 
\begin{eqnarray}
G_\epsilon(J_\epsilon) = \left\{ g \in U(n) \, \big| \, g^{\rm T} J
g = J \right\} \ . 
\end{eqnarray} 
Conversely, we can say that $J_\epsilon$ is an invariant tensor of $G_\epsilon(J_\epsilon)$.\\ 
{\bf Corollary 1:}
Arbitrary two elements $J,J' \in {\rm Inv}_\epsilon(n)$ are 
related to each other with appropriate unitary matrix $u$ as,
$J' = u \, J \, u^{\rm T}$ and corresponding group $G_\epsilon(J)$ and $G_\epsilon(J')$ are isomorphic to each other. {\tiny $\blacksquare$}\\
Therefore, from (\ref{eq:O}) and (\ref{eq:USp}) we find that $G_+(J_+)$ is isomorphic to $O(n)$ and $G_-(J_-)$ is isomorphic to $USp(n=2m)$.
\subsection{Diagonalization of the Vacuum Configuration}
{\bf Theorem 2-s:}
Let us consider an arbitrary $n$-by-$m$ $(n \le m)$  matrix $Q$ satisfying
\begin{eqnarray}
 QQ^\dagger =(Q Q^\dagger)^{\rm T}\ .
\end{eqnarray}
Then $Q$ is always decomposed as
\begin{eqnarray}
Q ~=~ O \left( 
\begin{array}{ccc|ccc} 
\lambda_1 & & & 0 & \cdots & 0 \\
 & \ddots & & \vdots & \ddots & \vdots \\
 & & \lambda_n & 0 & \cdots & 0 
\end{array}
\right) U\ ,
\end{eqnarray}
where $O\in SO(n)$ with $J={\bf 1}_n$ and $U\in U(m)$. \\
{\bf Proof 2-s:}
Since $QQ^\dagger$ is symmetric and Hermitian, $QQ^\dagger$ is a real symmetric matrix. Therefore it can be diagonalized as $QQ^\dagger=O\Lambda^2O^{\rm T}$ with $\Lambda={\rm diag(\lambda_1,\lambda_2,\cdots,\lambda_n)}$ with $\lambda_i\in {\mathbb R}_{\ge 0}$. {\tiny $\blacksquare$}

\noindent{\bf Theorem 2-a:}
Let us consider an arbitrary $2n$-by-$m$ $(2n \le m)$ matrix $Q$ satisfying
\begin{eqnarray}
 J QQ^\dagger =(Q Q^\dagger)^{\rm T}J\ ,
\end{eqnarray}
with $J= i\sigma_2 \otimes {\bf 1}_n$.
Then $Q$ can always be decomposed as
\begin{eqnarray}
Q =O \left( 
\begin{array}{ccc|ccc} 
 & & ~~& 0 & \cdots & 0 \\
 & \Lambda & & \vdots & \ddots & \vdots \\
 & & & 0 & \cdots & 0 
\end{array}
\right) U \ ,
\end{eqnarray}
where $O\in USp(2n)$ and $U\in U(m)$ and 
$\Lambda= {\bf 1}_2 \otimes {\rm diag(\lambda_1,\lambda_2,\cdots,\lambda_n)}$
with $\lambda_i\in {\mathbb R}_{\ge 0}$.
{\tiny $\blacksquare$}\\
{\bf Proof 2-a:}
The Hermitian positive semi-definite matrix  $QQ^\dagger$ is 
always diagonalized as
$QQ^\dagger =u \Lambda^2 u^\dagger$ 
with an appropriate unitary matrix $u\in U(2n)$.
Then the condition tells us that $X = u^{\rm T} J u$ commutes with
$\Lambda^2$, $[X,\Lambda^2]=0$. We can set $\Lambda$ to be positive
semi-definite, then  $[X,\Lambda]=0$.
Furthermore, we find $XX^\dagger={\bf 1}_{2n}$
and $X^{\rm T}=-X$. According to Theorem 1-a, thus, 
$X$ turns out to be $X=u^{\rm T}Ju=J$ by taking
an appropriate $u$. This means $u$ is an element of $USp(2n)$.
Here $\Lambda$ takes a form ${\bf 1}_2 \otimes \Lambda'$, 
since $[\Lambda,J]=0$. 
{\tiny $\blacksquare$}
\subsection{Diagonalization of a Non-Hermitian (Anti)symmetric Matrix}
\label{sec:DecompMatrix}
{\bf Theorem 3:}
An arbitrary $n$-by-$n$ (anti)symmetric matrix M, 
(that is, $M^{\rm T}=\epsilon M$) can be written in a block-diagonal form as
\begin{eqnarray}
M ~=~ u \, \left( 
\begin{array}{ccc} 
|\mu_{(1)}| J_{(1)} & & \\ 
& |\mu_{(2)}|J_{(2)} & \\
& & \ddots 
\end{array}
\right) u^{\rm T} \ , 
\end{eqnarray}
where $J_{(k)} \in {\rm Inv}_\epsilon(n_k)$ and $n = \sum_{k} n_k$.
{\tiny $\blacksquare$}\\
{\bf Proof 3:} 
$MM^\dagger$ is an Hermitian matrix and thus, can always be diagonalized as 
\begin{eqnarray}
MM^\dagger = u \, {\rm diag}\left(|\mu_{(1)}|^2{\bf 1}_{n_1},\,
|\mu_{(2)}|^2{\bf 1}_{n_2},\, \cdots \right) u^\dagger \ ,
\end{eqnarray}
with a unitary matrix $u$ and $|\mu_{(i)}|<|\mu_{(i+1)}|$.
Therefore, $\tilde M \equiv u^\dagger M u^*$ satisfies
\begin{eqnarray}
\tilde M \tilde M^\dagger &=& {\rm diag} \left( |\mu_{(1)}|^2{\bf 1}_{n_1},\, 
|\mu_{(2)}|^2{\bf 1}_{n_2},\, \cdots \right) \nonumber \\
&=& (\tilde M \tilde M^\dagger)^{\rm T}=\tilde M^\dagger \tilde M \ .
\label{eq:diagonal_M'}
\end{eqnarray} 
Note that $\tilde M^{\rm T}=\epsilon \tilde M$.
This equation means that $\tilde M$ is a normal matrix $[\tilde M, \tilde M^\dagger] = 0$ and can be diagonalized as
\begin{eqnarray}
\tilde M = \tilde u \, {\rm diag} \left(\mu_1,\, \mu_2,\, \cdots \right) \tilde u^\dagger\ ,
\end{eqnarray} 
with a unitary matrix $\tilde u$.
By substituting this form to Eq.({\ref{eq:diagonal_M'}}),
we find that
\begin{eqnarray}
 |\mu_{(1)}|^2=|\mu_1|^2=|\mu_2|^2=\cdots, \quad
 |\mu_{(2)}|^2=|\mu_{n_1+1}|^2=\cdots, \quad |\mu_{(3)}|^2=\cdots\ .
\end{eqnarray}
and $\tilde u$ should take a block-diagonal form 
as 
\begin{eqnarray}
\tilde u={\rm diag}(u_{(1)},u_{(2)},\cdots)\ ,
\end{eqnarray}
where $u_{(k)}$ is an $n_k$-by-$n_k$ unitary matrix.
Therefore, $\tilde M$ also takes block-diagonal form as
\begin{eqnarray}
\tilde M = {\rm diag} \left(
|\mu_{(1)}| J_{(1)},\, |\mu_{(2)}|J_{(2)},\, \cdots \right)\ .
\end{eqnarray}
{\tiny $\blacksquare$}\\
The meson field is always 'diagonalized' by fixing the flavor
symmetry.
Combining Theorem 1-s(1-a) with Theorem 3, 
we find the following corollaries.
\\
{\bf Corollary 3-s:}
An arbitrary symmetric matrix $M$ can be diagonalized
\begin{eqnarray}
 M=u\,m\, u^{\rm T},\quad m = {\rm diag}(|\mu_1|,\, |\mu_2|,\, \cdots)\ , 
\label{eq:diagonal}
\end{eqnarray}
with a unitary matrix $u$. {\tiny $\blacksquare$}\\
{\bf Corollary 3-a:}
An arbitrary anti-symmetric matrix $M$ can be diagonalized
\begin{eqnarray}
M = u \, m \, u^{\rm T},\quad m = 
\left(\begin{array}{cc}
0 & 1 \\
-1& 0
\end{array}\right) \otimes
{\rm diag}(|\mu_1|,\, |\mu_2|,\, \cdots)\ ,
\end{eqnarray}
with a unitary matrix $u$. {\tiny $\blacksquare$}\\
{\bf Corollary 3':}
An arbitrary $n$-by-$n$ 
(anti-)symmetric matrix $M$ can be decomposed as
\begin{eqnarray}
M = Q^{\rm T} J Q \ . 
\end{eqnarray}
where $Q$ is an $n$-by-$m$ matrix 
and $J \in {{\rm Inv}_\epsilon(m)}$
with $m = {\rm rank}(M)$. {\tiny $\blacksquare$}

The (anti)symmetric matrix $M$ breaks the $U(n)$ symmetry $M \rightarrow u M u^{\rm T}$ as
\begin{eqnarray}
U(n) \quad \rightarrow \quad \left\{ \begin{array}{cl} \hs{1} & U(n_0) \times 
O(n_1) \times O(n_2) \times \cdots \\ \hs{1} & U(n_0) \times USp(2m_1) \times USp(2m_2) \times \cdots \\ \end{array} \right. ,
\end{eqnarray}
where $n_0$ is a number of zero-eigenvalues of $M$.

\section{Non-trivial Uniqueness Proof \label{app:unique_proof}}
\label{appuniqueness}

In this section, we prove the uniqueness of the solution to Eq.\,(\ref{SO,USp_Xeq}). Here we consider the $SO(\NC)$ case. We can always write the $\NC$-by-$\NF$ matrix $Q$ as 
\beq
Q = \left(\hat Q,\, {\bf 0}\right){\cal U}, \hs{5} {\cal U} \in
U(\NF)\ ,
\eeq
up to $U(\NC)$ transformation which rotates the columns of the $\NC$-by-$\NC$ matrix of $\hat Q$. 
We can show that for $\hat M \equiv \hat Q^{\rm T} J \hat Q$
\begin{eqnarray}
{\rm rank} \, \hat M = \NC ~~~~~ \quad &\Leftrightarrow& \quad {\rm rank} \, \hat Q = \NC\ , \nn
{\rm rank} \, \hat M = \NC-1 \quad &\Rightarrow& \quad {\rm rank} \, \hat Q = \NC-1\ ,
\end{eqnarray}
since $ \det \hat M = \det J (\det \hat Q)^2$ and $\NC \ge {\rm rank} \, \hat Q \ge {\rm rank} \, \hat M$ is always satisfied. 

\subsection{Solution with ${\rm rank}\, M=\NC$} 
If the rank of $M \equiv Q^{\rm T} J Q$ is $\NC$, then $\hat M$ also has rank $\NC$. Therefore ${\rm rank} \, \hat Q = \NC$, namely $\hat Q$ is invertible and 
\begin{eqnarray}
U_Q \equiv \hat Q^{-1}\sqrt{\hat Q\hat Q^\dagger}\ ,
\end{eqnarray}
is a unitary matrix, $U_Q\in U(\NC)$.
In terms of this unitary matrix, we rewrite Eq.(\ref{SO,USp_Xeq}) as
\begin{eqnarray}
X&=&\sqrt{QQ^\dagger}e^{-V'}\sqrt{QQ^\dagger}=
U_Q^\dagger \hat Q^\dagger e^{-V'} \hat Q U_Q\ ,\nn
 X^2&=&\left(Q^{\rm T}J\sqrt{QQ^\dagger}\right)^\dagger 
Q^{\rm T}J\sqrt{QQ^\dagger}\nn
&=&U_Q^\dagger \hat Q^\dagger J^\dagger Q^*Q^{\rm T}J\hat Q U_Q
=U_Q^\dagger \hat M^\dagger \hat M U_Q\ . 
\end{eqnarray}
Since $\hat Q$ and $\hat M$ are invertible, we find 
a unique solution of $V'$
\begin{eqnarray}
 V'=\log\left(\hat Q\frac1{\sqrt{\hat M^\dagger \hat M}}\hat
 Q^\dagger\right)\ .
\end{eqnarray}
\subsection{Solution with ${\rm rank}\,M=\NC-1$}
In this case ${\rm rank} \, \hat Q = \NC-1$, we can use the $U(\NC)$ rotation so that the $\NC$-by-$\NC$ matrix $\hat Q$ takes the form
\beq
\hat Q = \left( 
\begin{array}{ccc|c} 
 & & & 0 \\ & \tilde Q & & \vdots \\ & & & 0 
\end{array} \right),
\eeq
where $\tilde Q$ is an $\NC$-by-($\NC-1$) matrix. 
We can introduce an $\NC$-component vector $p$ such that 
\begin{eqnarray}
 p^{\rm T}J\hat Q=p^{\rm T}J Q=0\ ,\quad p^{\rm T}J p=1\ ,
\end{eqnarray}
and the following $\NC$-by-$\NC$ matrix has the maximal rank
\begin{eqnarray}
R \equiv \left(\tilde Q, p\right) \in GL(\NC,{\mathbb C})\ . 
\end{eqnarray} 
Note that with a given $\tilde Q$, the column vector $p$ is uniquely determined up to sign.
Since $R$ is invertible,
$e^{V'}$ can be decomposed as
\begin{eqnarray}
 e^{V'}=R\left(
\begin{array}{cc}
 B& c\\c^\dagger & a
\end{array}
\right)R^\dagger\ .
\end{eqnarray} 
Here, $B$ is an $(\NC-1)$-by-$(\NC-1)$ 
Hermitian matrix and $a$ is a real parameter.
Eq.(\ref{SO,USp_eom_senza_lambda}) can be rewritten as
\begin{eqnarray}
 e^{V^{'\rm T}}JQQ^\dagger =Q^*Q^{\rm T}Je^{V'}\ .
\end{eqnarray}
Substituting the above decomposition and
 multiplying $R^{\rm T}J^*$ from the left 
and $J^\dagger R^*$ from the right, we find that
\begin{eqnarray}
B^{\rm T}\hat M=\hat M B, \quad c=0\ .
\end{eqnarray}
From the condition for $e^{V'} \in SO({\NC})^{\mathbb C}$,
we find the following equations 
\begin{eqnarray}
 a^2=1,\quad~~ \hat M^\dagger B^{\rm T}\hat M B={\bf 1}_{\NC-1}\ .
\end{eqnarray}
Note that we can say that $B$ and $a$ are positive definite since $c=0$.
Combining the above two equations, we obtain  
\begin{eqnarray}
B = \frac{1}{\sqrt{\hat M^\dagger\hat M}}\ ,\qquad a=1\ .
\end{eqnarray}
Therefore we finally find a unique solution
\begin{eqnarray}
 e^{V'} = \tilde Q \frac{1}{\sqrt{\hat M^\dagger \hat M}} \tilde Q^\dagger+p
 p^\dagger\ .
\end{eqnarray}
Note that $pp^\dagger$ is uniquely determined for a given $\tilde Q$, namely for a given $Q$.
Even if we could construct a similar solution for $V'$ 
in the case of ${\rm rank}\,M < \NC-1$,
it is obviously expected 
that a matrix corresponding to $pp^\dagger$ 
would not be unique.
These results exactly reflect the appearance of a partial Coulomb phase
in the case of ${\rm rank}\,M < \NC-1$.

\section{Deformed K\"ahler Potential for $USp(2\MC)$}
\label{appuspdeform}
The expansion of the deformed K\"ahler potential of
Eq.~(\ref{eq:Kuspdeform}) reads
\begin{align}
K_{USp,{\rm deformed}} =&\
\frac{1}{2}\sum_{i,j}\frac{1}{\mu'_i+\mu'_j}\left[1
+\frac{\varepsilon^2}{\mu'_i\mu'_j}\right]\phi_{ij}\phi_{ji}^\dag
\nn 
&-\frac{1}{2}\sum_{i,j,k}\frac{\mu_i}
{(\mu'_i+\mu'_j)(\mu'_i+\mu'_k)(\mu'_j+\mu'_k)}\left[1+
\varepsilon^2\frac{\mu'_i+\mu'_j+\mu'_k}{\mu'_i\mu'_j\mu'_k}\right]
\phi_{ij}\phi_{jk}^\dag(\phi J^\dag)_{ki} + {\rm c.c.} \nn
&+\sum_{i,j,k,l}X(\varepsilon)_{ijkl}
(\phi J^\dag)_{ij}(\phi J^\dag)_{jk}\phi_{kl}\phi_{li}^\dag 
+ {\rm c.c.} \nn
&+\frac{1}{2}\sum_{i,j,k,l}\frac{\mu_j\mu_l}{P'_{ijkl}}
\left[C_{ijkl}^{(1)'}
+\varepsilon^2\frac{C_{ijkl}^{(1)'}C_{ijkl}^{(2)'} -
  C_{ijkl}^{(3)'}}{C_{ijkl}^{(4)'}}
\right](\phi J^\dag)_{ij}\phi_{jk}\phi_{kl}^\dag(J \phi^\dag)_{li} \nn
&-\frac{1}{4}\sum_{i,j,k,l}
\left[\frac{C_{ijkl}^{(3)'}}{P'_{ijkl}}
+2\varepsilon^2\frac{C_{ijkl}^{(1)'}}{P'_{ijkl}}
+\varepsilon^4\frac{C_{ijkl}^{(1)'}C_{ijkl}^{(2)'}-C_{ijkl}^{(3)'}}
{C_{ijkl}^{(4)'}P'_{ijkl}}\right]
\phi_{ij}\phi_{jk}^\dag\phi_{kl}\phi_{li}^\dag \nn
&+ \textrm{K\"ahler trfs.} + \mathcal{O}(\phi^5) \ ,
\end{align}
where $\mu_i^{'2} \equiv \mu_i^2+\varepsilon^2$.
The resulting curvature is
\begin{align}
\left.R\right|_{\phi=0} =&\ 
-2\sum_i^{\MC}\frac{\mu_i^{'6} 
+ \varepsilon^2 7 \mu_i^{'4}
- \varepsilon^4 17 \mu_i^{'2}
- \varepsilon^6 7}{2\mu'_i(\varepsilon^2 + \mu_i^{'2})^3}\nn
&-2\sum_{i,j}^{\MC}\frac{\mu_i^{'4}\mu_j^{'4}}
{(\mu'_i+\mu'_j)(\varepsilon^2+\mu_i^{'2})(\varepsilon^2+\mu_j^{'2})
(\varepsilon^2+\mu'_i\mu'_j)^2} \nn
&+2\varepsilon^2\sum_{i,j}^{\MC}\frac{\mu_i^{'2}\mu_j^{'2}
(\mu_i^{'2}+\mu_j^{'2})}
{(\mu'_i+\mu'_j)(\varepsilon^2+\mu_i^{'2})(\varepsilon^2+\mu_j^{'2})
(\varepsilon^2+\mu'_i\mu'_j)^2} \nn
&-4\varepsilon^4\sum_{i,j}^{\MC}
\frac{\mu_i^{'2}\left(6\mu_i^{'2}+9\mu'_i\mu'_j+5\mu_j^{'2}\right)}
{(\mu'_i+\mu'_j)(\varepsilon^2+\mu_i^{'2})(\varepsilon^2+\mu_j^{'2})
(\varepsilon^2+\mu'_i\mu'_j)^2} \nonumber
\end{align}
\begin{align}
&-4\varepsilon^6\sum_{i,j}^{\MC}
\frac{\mu'_i\left(5\mu_i^{'2}+15\mu'_i\mu'_j+13\mu_j^{'2}\right)}
{\mu'_j(\mu'_i+\mu'_j)(\varepsilon^2+\mu_i^{'2})(\varepsilon^2+\mu_j^{'2})
(\varepsilon^2+\mu'_i\mu'_j)^2} \nn
&-2\varepsilon^8\sum_{i,j}^{\MC}
\frac{10\mu'_i+13\mu'_j}
{\mu'_j(\mu'_i+\mu'_j)(\varepsilon^2+\mu_i^{'2})(\varepsilon^2+\mu_j^{'2})
(\varepsilon^2+\mu'_i\mu'_j)^2} \nn
&+16\sum_{i,j,k}^{\MC}\frac{\mu_i^{'3}\mu_j^{'2}\mu_k^{'3}}
{(\mu'_i+\mu'_j)(\mu'_i+\mu'_k)(\mu'_j+\mu'_k)
(\varepsilon^2+\mu'_i\mu'_j)(\varepsilon^2+\mu'_i\mu'_k)
(\varepsilon^2+\mu'_j\mu'_k)} \nn
&+16\varepsilon^2\sum_{i,j,k}^{\MC}\frac{\mu_i^{'2}\mu_j^{'2}\mu_k^{'2}}
{(\mu'_i+\mu'_j)(\mu'_i+\mu'_k)(\mu'_j+\mu'_k)
(\varepsilon^2+\mu'_i\mu'_j)(\varepsilon^2+\mu'_i\mu'_k)
(\varepsilon^2+\mu'_j\mu'_k)} \nn
&+16\varepsilon^4\sum_{i,j,k}^{\MC}\frac{\mu'_i\mu'_k
\left(4\mu_i^{'2}\mu'_j+2\mu_i^{'2}\mu'_k+6\mu'_i\mu_j^{'2}+3\mu_j^{'3}
+4\mu'_i\mu'_j\mu'_k\right)}
{\mu'_j(\mu'_i+\mu'_j)(\mu'_i+\mu'_k)(\mu'_j+\mu'_k)
(\varepsilon^2+\mu'_i\mu'_j)(\varepsilon^2+\mu'_i\mu'_k)
(\varepsilon^2+\mu'_j\mu'_k)} \nn
&+16\varepsilon^6\sum_{i,j,k}^{\MC}
\frac{\mu_i^{'2}(\mu'_j+\mu'_k)+\mu'_j(\mu'_j+\mu'_k)^2
+\mu'_i(2\mu_j^{'2}+2\mu'_j\mu'_k+\mu_k^{'2})}
{\mu'_j(\mu'_i+\mu'_j)(\mu'_i+\mu'_k)(\mu'_j+\mu'_k)
(\varepsilon^2+\mu'_i\mu'_j)(\varepsilon^2+\mu'_i\mu'_k)
(\varepsilon^2+\mu'_j\mu'_k)} \ .
\end{align}

\end{document}